\documentclass[12pt]{iopart}

\usepackage{iopams}  

\usepackage{graphicx}

\usepackage[english]{babel}

\begin{document}

\title{The SUSY partners of the QES sextic potential revisited}

\author{Alonso Contreras-Astorga}
\address{CONAHCYT- Physics Department, Cinvestav, P.O. Box. 14-740, 07000, Mexico City, Mexico}
\ead{alonso.contreras@cinvestav.mx}

\author{A. M. Escobar-Ruiz}
\address{Departamento de F\'{i}sica, Universidad Aut\'onoma Metropolitana Unidad Iztapalapa, San Rafael Atlixco 186, 09340 Cd. Mx., M\'exico}
\ead{admau@xanum.uam.mx}

\author{Román Linares}
\address{Departamento de F\'{i}sica, Universidad Aut\'onoma Metropolitana Unidad Iztapalapa, San Rafael Atlixco 186, 09340 Cd. Mx., M\'exico}
\ead{lirr@xanum.uam.mx}




\begin{abstract}
In this paper, the SUSY partner Hamiltonians of the quasi-exactly solvable (QES) sextic potential $V^{\rm qes}(x) = \nu\, x^{6} +  2\, \nu\, \mu\,x^{4} +  \left[\mu^2-(4N+3)\nu  \right]\, x^{2}$, $N \in \mathbb{Z}^+$, are revisited from a Lie algebraic perspective. It is demonstrated that, in the variable $ \tau=x^2$, the underlying $\mathfrak{sl}_2(\mathbb{R})$ hidden algebra of $V^{\rm qes}(x)$ is inherited by its SUSY partner potential $V_1(x)$ only for $N=0$. At fixed $N>0$, the algebraic polynomial operator $h(x,\,\partial_x;\,N)$ that governs the $N$ exact eigenpolynomial solutions of $V_1$ is derived explicitly. These odd-parity solutions appear in the form of zero modes. The potential $V_1$ can be represented as the sum of a polynomial and rational parts. In particular, it is shown that the polynomial component is given by $V^{\rm qes}$ with a different non-integer (cohomology) parameter $N_1=N-\frac{3}{2}$.  A confluent second-order SUSY transformation is also implemented for a modified QES sextic potential possessing the energy reflection symmetry. By taking $N$ as a continuous real constant and using the Lagrange-mesh method, highly accurate values ($\sim 20$ s. d.) of the energy $E_n=E_n(N)$ in the interval $N \in [-1,3]$ are calculated for the three lowest states $n=0,1,2$ of the system. The critical value $N_c$ above which tunneling effects (instanton-like terms) can occur is obtained as well. At $N=0$, the non-algebraic sector of the spectrum of $V^{\rm qes}$ is described by means of compact physically relevant trial functions. These solutions allow us to determine the effects in accuracy when the first-order SUSY approach is applied on the level of approximate eigenfunctions.

\end{abstract}


\section{Introduction}

In quantum mechanics, quasi-exactly solvable (QES) systems are spectral problems, ${\cal H}\,\psi=E\,\psi$, for which it is possible to obtain in closed (analytical) form only a finite number of exact solutions while the remaining ones remain unknown. A systematic approach to quasi-exact solvability is purely algebraic \cite{Turbiner1988QuasiexactlysolvablePA, TURBINER1987181, U1988exact}. In this case, the Hamiltonian can be rewritten as a constant coefficient quadratic combination in the generators of a Lie algebra. The underlying hidden algebraic structure then leads to non-trivial dynamical features of these systems. For instance, one can mention the appearance of the so-called energy-reflection (ER) symmetry \cite{PhysRevA.59.1791} as well as the existence of a generating function for a set of orthogonal polynomials $P_n(E)$ in the energy variable $E$ \cite{10.1063/1.531373}. 

Interestingly, starting from a given QES model, the generation of new classes of QES problems was pointed out by means of supersymmetric quantum mechanics (SUSY) techniques \cite{Shifman1989NEWFI, Roy1991ONQS, ROY1989427, Gangopadhyaya1995}. A key element is the factorization method as described first in Dirac’s book \cite{Dirac1930-DIRTPO} and further developed in Ref. \cite{Infeld1951TheFM}, whereas complete reviews on supersymmetric quantum mechanics can be found in Ref. \cite{COOPER1995267,bagchi2000,Andrianov2004,David2010,Junker2019}. In particular, in Ref. \cite{Gangopadhyaya1995}, the SUSY framework is one of the three methods employed to derive new QES potentials. Even more, in the last two decades, much attention has been paid to quasi-solvable models from the point of view of ${\cal N}$-fold supersymmetry \cite{ANDRIANOV1993273,AOYAMA2001105} (see also \cite{TANAKA2003413} and references therein).

In most cases, the key ideas about QES models are illustrated taking as a prototype the sextic potential $V^{\rm qes}(x) =  \nu^{2}\,x^{6} +  2\,\nu\, \mu\, x^{4} +  (\mu^{2}\,-\,(4N+3)\nu) \ x^{2}$. This one-dimensional problem possesses a hidden $\mathfrak{sl}_2$ Lie algebra. If the parameter $N$ is a positive integer number, it is possible to find only $(N+1)$ exact analytical solutions. The exact ground state $\psi^{\rm qes}_0(x;N)$ can always be obtained explicitly. Therefore, the first-order SUSY partner Hamiltonian $\mathcal{H}_{1}$ with potential $V_1(x)$ can also be constructed for any positive integer value of $N$. A natural question arises, namely, \textit{Does the SUSY partner potential of $V^{\rm qes}$ possess a hidden Lie algebra?}. To the best of the authors' knowledge, such a relevant question has not been previously studied in the literature. In this respect, the situation is very different from previous works where the emphasis relies on finding methods for generating QES potentials. Here, we focus on the search for a hidden algebraic structure and special properties of the first-order partner Hamiltonian $\mathcal{H}_{1}$ and its solutions. 

Along these lines, we also explore the sextic QES system combining the SUSY framework and the variational method to analyze the connection, in the non-algebraic sector of the spectrum, between the bosonic and fermionic QES Hamiltonians. Since highly accurate approximate solutions are constructed, the description of the first-order SUSY mechanism at the level of approximate solutions is investigated in detail.

The goal of the present study is threefold. Firstly, for the lowest values of the parameter $N=0,1,2$, we will derive an algebraic polynomial operator $h(x,\,\partial_x;\,N)$. As a distinguished feature, the zero modes of $h(x,\,\partial_x;\,N)$ give the $N$ exact polynomial eigenfunctions of the SUSY partner potential $V_1(x)$ which, in general, is not a polynomial but a rational function in the variable $x^2$. Towards a Lie-algebraic characterization, the analytical properties of $V_1$ are clearly indicated. For $N>2$, the generic expression of $h(x,\,\partial_x;\,N)$ is presented as well. To overcome the 1-SUSY requirement of the explicit knowledge of the exact (analytical) ground state function $\psi^{\rm qes}_0(x;N)$ a confluent second-order SUSY transformation is applied on a QES sextic potential to generate isospectral QES models.

Secondly, for the three lowest states $n=0,1,2$ of the QES potential $V^{\rm qes}(x)$, highly accurate values of the energy $E_n=E_n(N)$ as a function of $N$ are displayed within the interval $N \in [-1,3]$. Thus, we investigate the system with $N$ being considered as a continuous real parameter. The numerical results are presented with $\sim 20$ significant digits; the corresponding calculations are performed using the Lagrange-mesh method. In particular, the critical value $N_c$ above which tunneling effects can occur is computed.  

Finally, for the case $N=0$ where solely the exact ground state of $V^{\rm qes}(x)$ is known, we generate approximate solutions for $V_1(x)$. They are constructed by means of SUSY techniques allied with the variational method. As a first step, compact variational trial functions $\psi_{\rm trial}$ for the excited states of $V^{\rm qes}$ are designed. They encode relevant physical properties of the system. Afterward, a SUSY transformation acting on these $\psi_{\rm trial}$ will produce approximate eigenfunctions for $V_1$. This simple idea allows us to determine the effects in accuracy when SUSY is implemented on the level of approximate eigenfunctions, another interesting practical aspect absent in the literature. 

\section{Generalities}
We consider the following one-dimensional spectral problem in non-relativistic quantum mechanics:
\begin{equation}
\label{Hp}
    {\cal H}\,\psi(x)\ = \ E\,\psi(x) \ ,   \qquad \quad \psi(x) \in {\cal L}^2 \ , 
\end{equation}
defined on the real line, $x \in (-\infty,\infty)$. The corresponding Hamiltonian operator for our specific problem is of the form:
\begin{equation}
\label{HQES}
    {\cal H} \ = \ -\frac{\hbar^2}{2\,m}\frac{d^2}{dx^2} \ + \ \frac{1}{2}\,V^{\rm qes}(x) \ , 
\end{equation}
where $m$ denotes the mass of the particle, and
\begin{equation}
\label{vqes}
V^{\rm qes}(x)\ = \ \nu^{2}\,x^{6}\ + \ 2\,\nu\, \mu\, x^{4}\ + \ [\mu^{2}\,-\,(4N+3)\nu] \ x^{2} \ ,
\end{equation}
is the quasi-exactly solvable sextic potential, here $\nu, \mu, N$ are real parameters. At $\nu=0$, the potential (\ref{vqes}) reduces to the exactly-solvable simple harmonic oscillator $V^{\rm es}(x)=\mu^2\,x^2$.
Hereafter, we will adopt atomic units $\hbar=1$, $m=1$.

\begin{figure}[t]
\centering
\includegraphics[width=10cm]{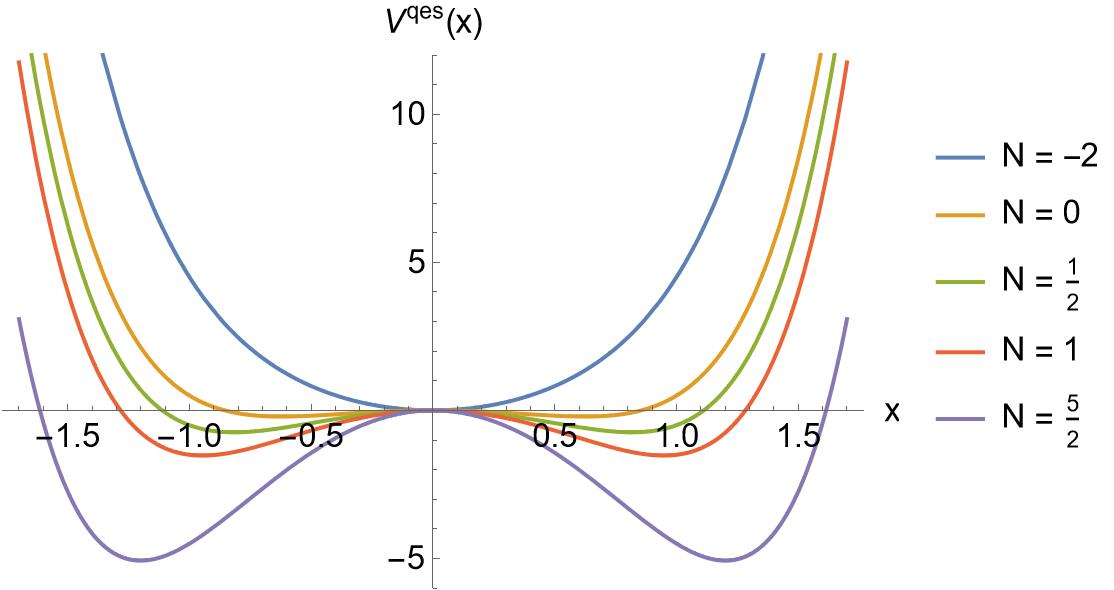}
\caption{At fixed $\mu=\nu=1$, we display the confining potential $V^{\rm qes}(x)$ in (\ref{vqes}) for different values of the parameter $N$. For $N>-\frac{1}{2}$ it develops two symmetric degenerate minima.}
\label{F1}
\end{figure}

Some remarks are in order:
\begin{itemize}
\item For $\nu \neq 0$ and arbitrary $N$ and $\mu$, $N$ not necessarily an integer number, the potential $V^{\rm qes}$ (\ref{vqes}) admits infinitely many bound states, see Fig. \ref{F1}. No scattering states occur. 

\item At fixed $\mu$ and $\,\nu>0$, there exists a special value $N=\frac{\mu ^2-3\, \nu }{4\, \nu }$ above which the potential (\ref{vqes}) develops two symmetric degenerate minima. Thus, tunneling effects (instanton-like terms) can take place. For large $N \rightarrow \infty$, these two minima are located at $x_{\pm}\approx \pm\,\sqrt{2}{(\frac{N}{3\,\nu})}^{1/4}$, respectively, and $V^{\rm qes}(x_{\pm}) \approx \frac{4}{9} N \left(3 \mu -2 \sqrt{3\,\nu\,N} \right)$.

\item The Hamiltonian (\ref{HQES}) is invariant under the parity symmetry $x \rightarrow -x$, which gives rise to alternating symmetric and anti-symmetric bound states.

\item At $\mu = 0$, $N$ and $\nu \neq 0$  arbitrary, the energy reflection (ER) symmetry in (\ref{HQES}) emerges \cite{PhysRevA.59.1791}. For instance, the wave functions of the ER symmetric levels are connected to each other by analytic continuation $x \rightarrow i\,x$, $\psi_E \rightarrow \psi_{-E}$.

\item Formally, the most general eigenfunction of (\ref{HQES}) reads
  \begin{eqnarray} \label{psiheun}
 \hspace{-1cm} \psi(x) \ &  = & \hspace{-0cm} \ e^{\frac{1}{4} \left(\nu \, x^4+2 \,\mu\,  x^2\right)}\times \bigg(\,c_1\, w\left[\frac{1}{4} (-2 \,E-\mu ),\nu  \left(N+\frac{3}{2}\right),\frac{1}{2},\mu ,\nu ,x^2\right]  \nonumber \\ 
    & ~ &\ + \ c_2 \,x \,w\left[\frac{1}{4} (-2 \,E-3 \mu ),\nu  (N+2),\frac{3}{2},\mu ,\nu ,x^2\right]\, \bigg) \ ,
 \end{eqnarray}
here $c_1$, $c_2$ are constants of integration, and $w(z)=w(q,\alpha ,\gamma ,\delta ,\epsilon ,z)$ satisfies the bi-confluent Heun differential equation $z\, w''+w' \,\left( z^2 \,\epsilon +\delta\,  z+\gamma\right)\,+\,w\, (\alpha \, z-q)=0$. We will focus on the case where $\psi(x)$ is normalizable, and one of its factors is a polynomial function in the $x-$variable. Other solutions are also interesting; see for example \cite{Levai2019} where a sextic potential with a centrifugal barrier is considered.
    \item The Hamiltonian (\ref{HQES}) possesses a hidden $\mathfrak{sl}_2(\mathbb{R})$ Lie algebra \cite{Turbiner1988QuasiexactlysolvablePA}. In the particular case when $N$ is a positive integer number, $N\in \mathbb{Z}^+$, this algebra admits a finite-dimensional irreducible representation and, thus, one can find algebraically only $(N+1)$ exact even-parity eigenfunctions and their corresponding energies explicitly.

\item As stated in \cite{Kozçaz}, for a positive integer number $N$, the system (\ref{vqes}) is intimately related to the exact cancellation of real and complex nonperturbative saddles to all orders in the semiclassical expansion.

\item Finally, at $N=-j-\frac{1}{2}$, $j$ a positive integer, we arrive to the sextic $\cal{P\,T}-$symmetric QES potential \cite{Bender2005}. 
    
\end{itemize}

\section{The hidden $\mathfrak{sl}_2$ Lie algebra of $V^{\rm qes}(x)$}

We begin with the well-established result that the Hamiltonian $\cal H$ in (\ref{HQES}) can be transformed into a suitable $\mathfrak{sl}_2$ Lie algebraic operator \cite{Turbiner1988QuasiexactlysolvablePA},\cite{TURBINER20161}. To this end, the gauge factor: 
\begin{equation}
    \Gamma(x) \ = \  \exp{[-\frac{\nu}{4}\,x^{4}-\frac{\mu}{2}\,x^{2}]}\ ,
\end{equation}
is introduced. In the $\mathbb{Z}_2$-invariant variable  $z \ = \ x^2$, 
one can construct the gauge-rotated Hamiltonian
\begin{equation}
\label{oph}
    h \ \equiv \   \Gamma^{-1}
    \,{\cal H} \, \Gamma \ = \  - 2\,z\,\frac{d^2}{dz^2} \ + \ (2\,\nu\,z^2+2\,\mu\,z-1)\frac{d}{dz} \ - \ 2\,N\,\nu\,z \ + \ \frac{\mu}{2} \ ,
\end{equation}
which can be rewritten as a constant coefficient quadratic combination in terms of the $\mathfrak{sl}_2$ generators \cite{Turbiner1988QuasiexactlysolvablePA}
\begin{equation}
\label{gener}
 {\cal J}^+_N(\tau) \ = \ \tau^2\,\frac{d}{d\tau}  \ - \ N\,\tau, \quad  
{\cal J}^0_N(\tau) \ = \ \tau\,\frac{d}{d\tau}  \ - \ \frac{N}{2}, \quad 
{\cal J}^-_N(\tau) \ = \ \frac{d}{d\tau}  \ .
\end{equation}
Explicitly,
\begin{equation}
\label{ophh}
    h \ = \  -2\,{\cal J}^0_N\,{\cal J}^-_N \ + \ 2\,\nu\,{\cal J}^+_N \ + \ 2\,\mu\,{\cal J}^0_N \ - \ (N+1)\,{\cal J}^-_N \ + \ \mu\,\big(N \, + \, \frac{1}{2}\big)\ .
\end{equation}
Moreover, if the parameter $N$ takes positive integer values, then the spectral problem 
\begin{equation}
h\,P(z)\ =\  E\,P(z) \ ,    
\end{equation}
possesses ($N+1$)- exact polynomial eigenfunctions $P_j$ with $j=0,1,2,\ldots,N$. In this case, the ($N+1$)- exact solutions of the original Hamiltonian $\cal H$ in (\ref{Hp}) take the form
\begin{equation}
    \psi_j(z)\ = \ P_j(z)\times \exp{[-\frac{\nu}{4}\,z^{2}-\frac{\mu}{2}\,z]} \ , \qquad z=x^2 \ .
\end{equation}
All of them correspond to even-parity states, including the ground state, possessing an even number of nodes. It is worth mentioning that odd-parity exact solutions of $\cal H$ (\ref{Hp}) appear for positive half-integer values of $N=\frac{1}{2},\frac{3}{2}\,\ldots$. They correspond to excited states and can be associated with the existence of the same hidden $\mathfrak{sl}_2$ algebra. However, the basic object we will use later in first-order supersymmetric quantum mechanics is precisely the ground state solution. Therefore, the present study does not consider such odd-parity exact states of $\cal H$ (\ref{Hp}) except in the part concerning the confluent second-order SUSY transformation. 

\subsection{Exact even-parity polynomial solutions: cases $N=0,1,2,3$}

This section reviews explicit results for the lowest positive integer values of $N$. They are well-known in the literature and will be used later in the present consideration. For convenience, we take $\nu=1$ and $\mu=1$. The QES sextic potential reads
\begin{equation}\label{Vred}
V^{\rm qes}(x,\,N)\ = \ x^{6}\ + \ 2\, x^{4}\ - \ 2\,(2N+1) \, x^{2} \ ,
\end{equation}
see Fig. \ref{F1}. The corresponding exact solutions of the Schrödinger equation are the following: 
\begin{itemize}
    \item At $N=0$, one exact solution occurs only. It is given by the nodeless function:
\begin{equation}
\label{psi0N0}
    \psi^{(N=0)}_0 \ = \ \exp{[-\frac{1}{4}\,x^{4}-\frac{1}{2}\,x^{2}]} \ ,
\end{equation}
with energy
\begin{equation}
    E^{(N=0)}_0 \ = \ \frac{1}{2} \ .
\end{equation}

It corresponds to the exact ground state of the system. Accordingly, the polynomial factor $P^{(N=0)}_0\equiv 1$ is an eigenfunction of the $\mathfrak{sl}_2$ Lie-algebraic operator $h$ (\ref{ophh}).

    \item  At $N=1$, solely two exact solutions appear:

\begin{equation}
    \psi^{(N=1)}_0 \ = \  (2\,x^2+\sqrt{3}+1)\,\exp{[-\frac{1}{4}\,x^{4}-\frac{1}{2}\,x^{2}]} \ ,
\end{equation}
with energy
\begin{equation}
    E^{(N=1)}_0 \ = \ \frac{3}{2}-\sqrt{3} \ ,
\end{equation}
and
\begin{equation}
    \psi^{(N=1)}_2 \ = \  (2\,x^2-\sqrt{3}+1)\,\exp{[-\frac{1}{4}\,x^{4}-\frac{1}{2}\,x^{2}]} \ ,
\end{equation}
here
\begin{equation}
    E^{(N=1)}_2 \ = \ \frac{3}{2}+\sqrt{3} \ .
\end{equation}

They correspond to the ground and second excited state, respectively. The polynomial factors $P^{(N=1)}_0=2\,x^2+\sqrt{3}+1$ and $P^{(N=1)}_2 \ = \  2\,x^2-\sqrt{3}+1$ are exact eigenfunctions of the Lie-algebraic operator $h$ (\ref{ophh}).

    \item At $N=2$, following the general theory, three exact analytical solutions occur only:

\begin{equation}
    \psi^{(N=2)}_0 \ = \  (2\, x^4+6\, x^2+3)\,\exp{[-\frac{1}{4}\,x^{4}-\frac{1}{2}\,x^{2}]} \ ,
\end{equation}
with energy
\begin{equation}
    E^{(N=2)}_0 \ = \ -\frac{3}{2} \ .
\end{equation}

Next, 
\begin{equation}
    \psi^{(N=2)}_2 \ = \  (2 \,x^4+2 \sqrt{2}\, x^2-\sqrt{2}-1)\,\exp{[-\frac{1}{4}\,x^{4}-\frac{1}{2}\,x^{2}]} \ ,
\end{equation}
where
\begin{equation}
    E^{(N=2)}_2 \ = \ -2 \sqrt{2}+\frac{9}{2} \ ,
\end{equation}
and 
\begin{equation}
    \psi^{(N=2)}_4 \ = \  (2\, x^4-2 \sqrt{2} \,x^2+\sqrt{2}-1)\,\exp{[-\frac{1}{4}\,x^{4}-\frac{1}{2}\,x^{2}]} \ ,
\end{equation}
with eigenvalue
\begin{equation}
    E^{(N=2)}_4 \ = \ 2 \sqrt{2}+\frac{9}{2} \ .
\end{equation}

They correspond to the ground, second, and fourth excited states. Thus, the corresponding factors $P^{(N=2)}_0=2\, x^4+6\, x^2+3$, $P^{(N=2)}_2=2 \,x^4+2 \sqrt{2}\, x^2-\sqrt{2}-1$ and $P^{(N=2)}_4=2\, x^4-2 \sqrt{2} \,x^2+\sqrt{2}-1$ are also eigenpolynomials of the Lie-algebraic operator $h$ (\ref{ophh}).
    \item At $N=3$, four exact eigenfunctions exist:
\begin{eqnarray}
  \hspace{-1cm}  \psi^{(N=3)}_k \ & = & \ \left( 
        384 \,x^6 \  + \ 96 (13 - 2 E_k) \,x^4 \ + \ 12\, (-3 + 4 (-11 + E_k) E_k)\, x^2 \right. \nonumber     \\ 
     & &  \left. -1263 - 22 E_k + 108 E_k^2 - 8 E_k^3  
       \,\right)
\,\exp{[-\frac{1}{4}\,x^{4}-\frac{1}{2}\,x^{2}]}
 \ ,    
\end{eqnarray}
$k=1,2,3,4$, being $E_k$ the $k$th-root of the quartic equation
\begin{equation}
    16\,E^4 \ - \ 224\,E^3 \ + \ 56\, E^2 \ + \ 3560\,E-1191 \ = \ 0 \ ,
\end{equation}
respectively.
\end{itemize}

\subsection{QES sextic potential with energy reflection symmetry: examples with $N=0,1$} \label{ER symmetry}

As previously mentioned in the introduction, at $\mu=0$, the Hamiltonian (\ref{HQES}) with potential $V^{\rm qes}(x)$ as in (\ref{vqes}) possesses the energy reflection (ER) symmetry $E \leftrightarrow -E $ \cite{PhysRevA.59.1791}. For the potentials with ER symmetry, we study a more general QES potential, the two-parametric sextic potential $V^{\rm qes}_{ER}(x)$: 
\begin{equation} \label{ER potential}
    V^{\rm qes}_{ER}(x)\ = \ \nu^2 x^6 \ + \ \nu\,(4N+2\kappa+3) \,x^2\ , 
\end{equation}
where $\kappa=0,1$ indicates the parity of the algebraic solutions. The case $\kappa=0$ corresponds with the potential $V^{\rm qes}(x)$ in (\ref{vqes}) at $\mu=0$.

Let us revisit in detail concrete solutions using the most straightforward examples, namely $N=0$ and $N=1$. They will be exploited later in the confluent SUSY algorithm, a degenerate case of the second-order supersymmetric quantum mechanics:
\begin{itemize}
    \item At $N=0$, the potential, the single eigenfunction, and the corresponding eigenvalue are 
    \begin{equation}\label{ERN0}
    \hspace{-3cm} V^{\rm qes}_{ER}(x)= \nu^2 x^6\, -\, \nu\, (2\kappa +3) \,x^2\,,  \quad \psi_{ER}^{(N=0)}(x) = x^\kappa \, \exp \left(-\frac{\nu}{4} x^4 \right) \ ,  \quad E_{ER}^{(N=0)} = 0\,,  
    \end{equation}
    where $\nu > 0$ and $E_{ER}^{(N=0)} $ is the energy of the ground ($\kappa=0$) and first-excited ($\kappa=1$) state $\psi_{ER}^{(N=0)}$, respectively.     
    
    \item If $N=1$, there are two exact analytical solutions for each value of $\kappa$, the potential, the analytic eigenfunctions, and its energies are:
    \begin{eqnarray} \label{ER case}
     V^{\rm qes}_{ER}(x)\ &=& \  \nu^2 \,x^6\ -\  \nu\, ( 2\,\kappa +7)\, x^2, \nonumber \\
     \psi_{ER,\pm}^{(N=1)}(x) \ &=& \ x^k\,\big(\,2 \,\nu \,x^2 \mp \sqrt{2\,\nu\,(1+2\,\kappa)}\,\big)\, \exp \left(-\frac{\nu}{4} x^4 \right), \nonumber \\  
    E_{ER,\pm}^{(N=1)}  \ &=& \ \pm \sqrt{2\,\nu\,(1+2\,\kappa) }. 
    \end{eqnarray}
When $\kappa=0$, the previous expressions correspond to the ground ($-$) and second excited ($+$) states whereas at $\kappa=1$ they describe the first ($-$) and third excited ($+$) state, respectively.
    
\end{itemize}

\section{\emph{Exact}- and WKB-numerical solutions, case $N=0$  }

For any value of $N$, by means of the user-friendly LagrangeMesh Mathematica Package \cite{JCR} (LMMP), one can easily determine the \textit{exact}-numerical eigenfunctions and eigenvalues. It is worth mentioning that the underlying Lagrange Mesh Method is an approximate variational method simplified by a Gauss quadrature associated with a certain mesh; for further details, see Ref. \cite{JCR}. The high efficiency and direct control of the involved accuracy in arithmetic manipulations (and final results) allow us to obtain highly accurate energies using a personal laptop. For instance, an excessive value with 30 correct significant digits can be achieved in short CPU times. A similar situation occurs for the corresponding wave functions. Therefore, in practice, we will refer to these numerical results as exact solutions, dropping the term numerical.

Just for $N=0$, we present in Table \ref{Tabenergies} the energies $E_n$ and the expectation value $\langle\,x^2\,\rangle$ for the lowest ten bound states $n=0,1,2,\ldots,9$,  explicitly. The calculations used a Hermite mesh with 800 mesh points for the energy. A comparison with WKB results is indicated as well.

\begin{table}[h]
\centering
\caption{Case $N=0$: exact numerical solutions $E_{\rm exact}$ for the sextic potential $\frac{1}{2}V^{\rm qes}(x,\,N=0)$ defined by (\ref{Vred}). Both the energy and the corresponding expectation value $\langle\,x^2\,\rangle$ were obtained using the LagrangeMesh Mathematica Package \cite{JCR}. Results are displayed in atomic units. For comparison, the WKB calculations $E_{WKB}$ are presented as well. Here $\Delta E\equiv \frac{E_{WKB}-E_{\rm exact}}{E_{\rm exact}}$.}
\label{Tabenergies}
\begin{tabular}{c| c | c |c |c}
\hline
\hline
\hspace{0.2cm}  $n$  \hspace{0.4cm}&  $E_{\rm exact}$ & $E_{WKB}$ & \hspace{0.2cm} $\Delta E [\%]$ \hspace{0.2cm} &\hspace{0.1cm } $\langle\,x^2\,\rangle$\\ \hline
0     & \hspace{0.2cm } 0.50000000000000000000  \hspace{0.2cm } & 0.56369894 & 0.127 &\  0.2896023863  \\
1     & 2.18650052957281497982 & 2.49875313 & 0.142 & \ 0.6490041219 \\
2     & 4.87181666510578189419 & 5.22318989 & 0.072 & \ 0.8028757103 \\
3     & 8.13095355822955323057 & 8.52449556 & 0.048 & \ 0.9587575624\\
4     & 11.874846994114710710  & 12.3031631 & 0.036 &\ 1.0946098649\\
5     & 16.039784895786315299  & 16.4981169 & 0.028 & \ 1.2170135056 \\
6     & 20.581916287588558795  & 21.0669696 & 0.023 & \ 1.3293330894 \\
7     & 25.468762870933097825  & 25.9780930 & 0.019 & \ 1.4337122075 \\
8     & 30.675013729847648104  & 31.2067221 & 0.017 & \ 1.5316298628 \\
9    & 36.180224949476151095  & 36.7327857 & 0.015 &\ 1.6241563887 \\
\hline
\hline
\end{tabular}
\end{table}

In Table \ref{Tabenergies2}, the energy $E=E(N)$ as a function of the cohomology parameter $N$ is depicted for the three lowest states $n=0,1,2$. If $N>-\frac{1}{2}$, the potential $V^{\rm qes}(x,\,N)$ (\ref{Vred}) has a maximum located at $x=0$, and it corresponds to a zero energy value. It implies that for negative energies $E<0$, the dynamics takes place in a classically forbidden region. At fixed $n$, the energy $E_n(N)$ is a smooth decreasing function of the parameter $N$. In particular, there exists a critical value $N=N_c$ such that the ground state energy $E_0(N_c)$ of the system vanishes. Accordingly, for $N>N_c$, the appearance of tunneling effects (exponentially small instanton-like terms) are expected. Interestingly, if $N$ is a positive (half)integer number, the corresponding instanton-like terms are absent in the exact analytical solutions! This puzzle was resolved in \cite{Kozçaz} by including complex saddles. It is an interesting open question whether this is valid for non-positive (half)integer values of $N$.

\begin{table}[h]
\centering
\caption{Sextic potential $\frac{1}{2}V^{\rm qes}(x,\,N)$ defined by (\ref{Vred}): the energy $E_n=E_n(N)$ in atomic units, as a function of $N$, for the three lowest states $n=0,1,2$. As for the critical value (see text) we found $N_c \approx 0.73295312615213043$. The results were obtained using the LagrangeMesh Mathematica Package\cite{JCR}.}
\label{Tabenergies2}
{\footnotesize \begin{tabular}{c| c | c |c }
\hline
\hline
\hspace{0.2cm}  $N$  \hspace{0.4cm}&  $E_{0,\,\rm exact}$ & $E_{1,\,\rm exact}$ & \hspace{0.2cm} $E_{2,\,\rm exact}$ \\ \hline
-1     &  0.989580605436050838998 & 
          3.360990709529042484024 & 6.413007989851041421032  \\ 
 $-\frac{3}{4}$     & 0.881159828282351698813 & 3.088336298693491389507 &  6.040643954210947777719   \\
 $-\frac{1}{2}$     & 0.764532033014503629516 & 2.802626174338946321036 &  5.659402903419812200032   \\
$-\frac{1}{4}$     &  0.638138724545526132477 & 2.502513714945104658141 &  5.269586956813807724855   \\
$-\frac{1}{8}$     & 0.570681019914247060253 & 2.346594455340202692157 &  5.071641780887543784212   \\
$0$     & 0.500000000000000000000 & 2.186500529572814979823 &  4.871816665105781894194  \\
$\frac{1}{8}$     & 0.425761495459989697814 & 2.022020182796715324855 &  4.670258698594960339359  \\
$\frac{1}{4}$     & 0.347587607715820659135 & 1.852931987632204372661 & 4.467151413879439497334    \\
$\frac{1}{2}$     & 0.177671681805842046340 & 1.500000000000000000000 &  4.057235461202041394422   \\
$N_c $     & $9.2 \times 10^{-12}$ & 1.151993714577925631588 & 3.672629779631384060651  \\
$\frac{3}{4}$     & -0.0138436436130964772246 & 1.125756241203857723531 &  3.644453673510100233335   \\
$1$    & -0.2320508075688772935275 & 0.7281390966952123635756 &  3.232050807568877293527   \\
$2$     & -1.5000000000000000000000 & -1.1383762435615163784847 &  1.671572875253809902397   \\
$3$    &  -3.6166170356875860239609  & -3.5323497843209945952738 &  0.335095120779029553927   \\
\hline
\hline
\end{tabular}}
\end{table}

\section{QES potentials via supersymmetric quantum mechanics}

\subsection{First-order supersymmetry} \label{section 1susy}

We start from a certain Schrodinger operator $\mathcal{H}_{0}=-\frac{1}{2} \frac{d^2}{dx^2} \,+\, V_0(x)$, where some of its eigenfunctions ${\psi_n}(x)$ and eigenenergies $E_n$ are known explicitly. Then we suppose the existence of a first-order differential operator $A_1^+$ such that it intertwines $\mathcal{H}_{0}$ with a new Hamiltonian $\mathcal{H}_{1}=-\frac{1}{2} \frac{d^2}{dx^2} + V_1(x)$ in the following way
\begin{equation}
\label{intrel}
    \mathcal{H}_{1} \,A_1^+ \ = \ A_1^+\, \mathcal{H}_{0}\ ,
\end{equation}
where
\begin{equation}
\label{OpA1}
    A_1^+\ = \ \frac{1}{\sqrt{2}} \left(- \frac{d}{dx} \,+ \,\frac{u'(x)}{u(x)} \right) \ .
\end{equation}
In the last expression, $u(x)$ is called \emph{seed solution}, and it solves the equation $\mathcal{H}_{0}\, u = \epsilon\, u$, where $\epsilon$ is a constant known as \emph{factorization energy}. We use the notation $u'(x)=d u(x)/dx$. The seed solution $u$ is not necessarily a physical solution $\psi$ of $\mathcal{H}_{0}$. In the literature  $\alpha(x) \equiv u'(x)/ u(x)= (\ln u(x))'$ is called the \emph{superpotential}. From the intertwining relation (\ref{intrel}) it follows the expression 
\begin{equation}
    V_1\ = \ V_0\,-\,(\ln u)'' \label{V1 susy} \ .
\end{equation}
It is said that $V_1$ is the \emph{SUSY partner} of $V_0$. Moreover, from the equation above, the importance of avoiding zeroes of $u$ is clear, so the SUSY partner potential $V_1$ is regular in the same domain of $V_0$. This imposes the condition $\epsilon \leq E_0$. 

If we introduce the adjoint operator of $A^+_1$,  
\begin{equation}
    A_1\ = \ \frac{1}{\sqrt{2}} \left( \frac{d}{dx} \,+ \,\frac{u'(x)}{u(x)} \right) \ ,
\end{equation}
we can show by direct substitution that the operators $A_1$ and $A_1^+$ factorize the Hamiltonians as $\mathcal{H}_0 = A_1 \ A_1^+ + \epsilon $ and $\mathcal{H}_1 = A_1^+ \ A_1 + \epsilon$. 

Also, from the intertwining relation and the factorization of $\mathcal{H}_0$, it can be shown that if $\psi_k$ is an eigenfunction of $\mathcal{H}_0$ with eigenvalue $E_k$ then 
\begin{equation}\label{solutions 1susy}
    \phi_k = \frac{1}{\sqrt{E_k- \epsilon}} A_1^+ \psi_k \ ,
\end{equation}
is an eigenfunction of $\mathcal{H}_1$ with the same eigenvalue $E_k$. The factor $1/\sqrt{E_k-\epsilon}$ comes from the normalization condition.

\textbf{Remark} \textit{The subindex $k$ in $\psi_k$ indicates the energy level, i.e., $\psi_k$ is the $k$-th excited state  of $\mathcal{H}_0$. In contrast, in the transformed wavefunction $\phi_k$ of $\mathcal{H}_1$, it is not necessarily the $k$-th excited state, as we will show below.}

\noindent From the factorization of $\mathcal{H}_1$, we can see that the function annihilated by $A_1$ is also a solution of the eigenvalue equation of $\mathcal{H}_1$ with eigenvalue $\epsilon$. This function is known as the \emph{missing state} and its expression is 
\begin{equation} \label{missing state 1susy}
\phi_\epsilon \ \propto \ \frac{1}{u}\ . 
\end{equation}
If $\phi_\epsilon$ fulfills the boundary conditions, $\epsilon$ belongs to the spectrum of $\mathcal{H}_1$, otherwise it does not. In general, if $\epsilon_0 = E_0$ and $u(x)= \psi_0$ then Sp$(\mathcal{H}_1)$ differs from Sp$(\mathcal{H}_0)$ only on the ground state energy.

Below, we present the supersymmetric partner potential $V_1(x,N)$ corresponding to $V_0=\frac{1}{2}V^{\rm qes}(x,N)$ for the lowest cases $N=0,1,2,3$.

\subsection{1-SUSY partner potential $V_1$ with $N=0$}

We take $V_0=\frac{1}{2}V^{\rm qes}(x,N=0)$. 
In this case, we solely know the exact ground state solution $\psi_0^{(N=0)}$ with energy $E_0 = 1/2$. No exact analytical solutions for the excited states of $\mathcal{H}_{0}$ are known. Then, we can only choose $u=\psi_0^{(N=0)}$. Using (\ref{V1 susy}), we immediately obtain the SUSY partner potential 
\begin{equation}
\label{vn0}
   V_1(x,N=0) \ = \  \frac{x^6}{2}\ + \ x^4 \ + \ 2 \,x^2 \ + \ 1 \ ,
\end{equation}
which is in complete agreement with the general Eq. (31) in \cite{Gangopadhyaya1995}.

Up to an additive constant, $V_1$ (\ref{vn0}) coincides with the sextic potential $\frac{1}{2}V^{\rm qes}(x,N=-3/2)$, see (\ref{Vred}). Thus, just like $V_0$, it is an analytic function with no poles on the real (and complex) domain. Remarkably, the SUSY partner Hamiltonian $\mathcal{H}_{1}$ with potential (\ref{vn0}) still possesses a hidden $\mathfrak{sl}_2$ Lie algebraic structure but with a negative value of parameter $N$. The trivial SUSY solution of $\mathcal{H}_{1}\, \phi = E \,\phi$, namely, $\phi = 1/u$ is non-square integrable. 
\begin{figure}[t]
\centering
\includegraphics[width=7cm]{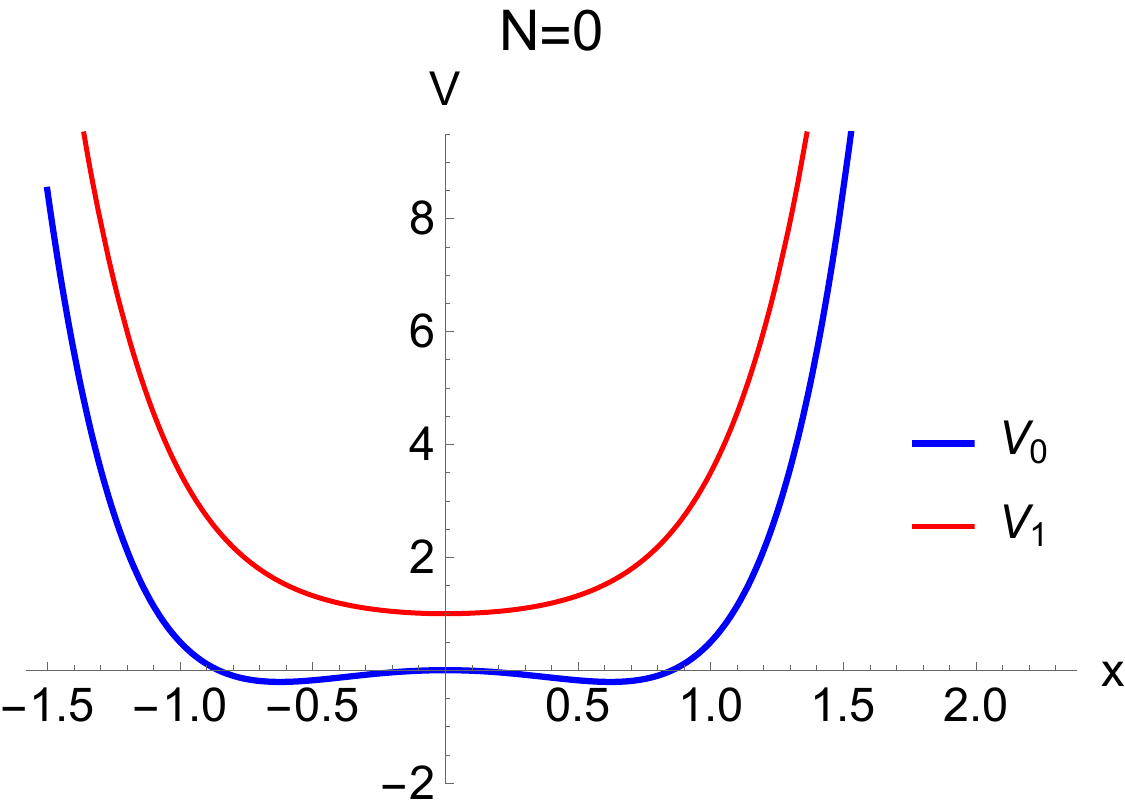}
\caption{For fixed $\mu=\nu=1$ and $N=0$, the confining potential $V_0=V^{\mathrm{qes}}(x)/2$ defined by (\ref{vqes}) (blue curve) and its SUSY partner $V_1(x)$ as in (\ref{vn0}). }
\end{figure}

\subsection{1-SUSY partner potential $V_1$ with $N=1$}
Here $V_0=\frac{1}{2}V^{\rm qes}(x,N=1)$, for which we know two exact analytical solutions, namely $\psi_0^{(N=1)}$ and $\psi_2^{(N=1)}$. Substituting $u=\psi_0$ in (\ref{V1 susy}) we obtain
\begin{eqnarray}\label{V1 qes}
    V_1(x,N=1)  \ & = &  \ \frac{x^6}{2}\ + \ x^4 \ + \ 1 \ - \ \frac{4 \left(-2 x^2+\sqrt{3}+1\right)}{\left[P^{(N=1)}_0\right]^2} \nonumber 
\\  
& = & \ \frac{1}{2} V^{\rm qes}(x,N=-1/2)\  + \ 1 \ - \ \frac{4 \left(-2 x^2+\sqrt{3}+1\right)}{\left[P^{(N=1)}_0\right]^2}\ ,
\end{eqnarray}
where $P^{(N=1)}_0=2\, x^2+\sqrt{3}+1$ vanishes only in the complex domain. The SUSY transformation retrieves a rational extension of $V^{\rm qes}$. For the above potential $V_1$, we know only a single exact solution (the first excited state)
\begin{equation}
  \phi_2 \ = \ A_1^+ \psi_2 \ = \  \frac{x}{P^{(N=1)}_0}\,   \exp \left(-\frac{1}{4} x^4-\frac{1}{2} x^2\right) \ ,
\end{equation}
with energy $E_2^{(N=1)}= \frac{3}{2}+\sqrt{3}$. It is very common that the SUSY transformation changes the parameters of a potential and adds a finite term. In this case, such a finite part is a rational function in the variable $x^2$, with complex poles given by the complex zeros of $P^{(N=1)}_0$. 

\begin{figure}[t]
\centering
\includegraphics[width=7cm]{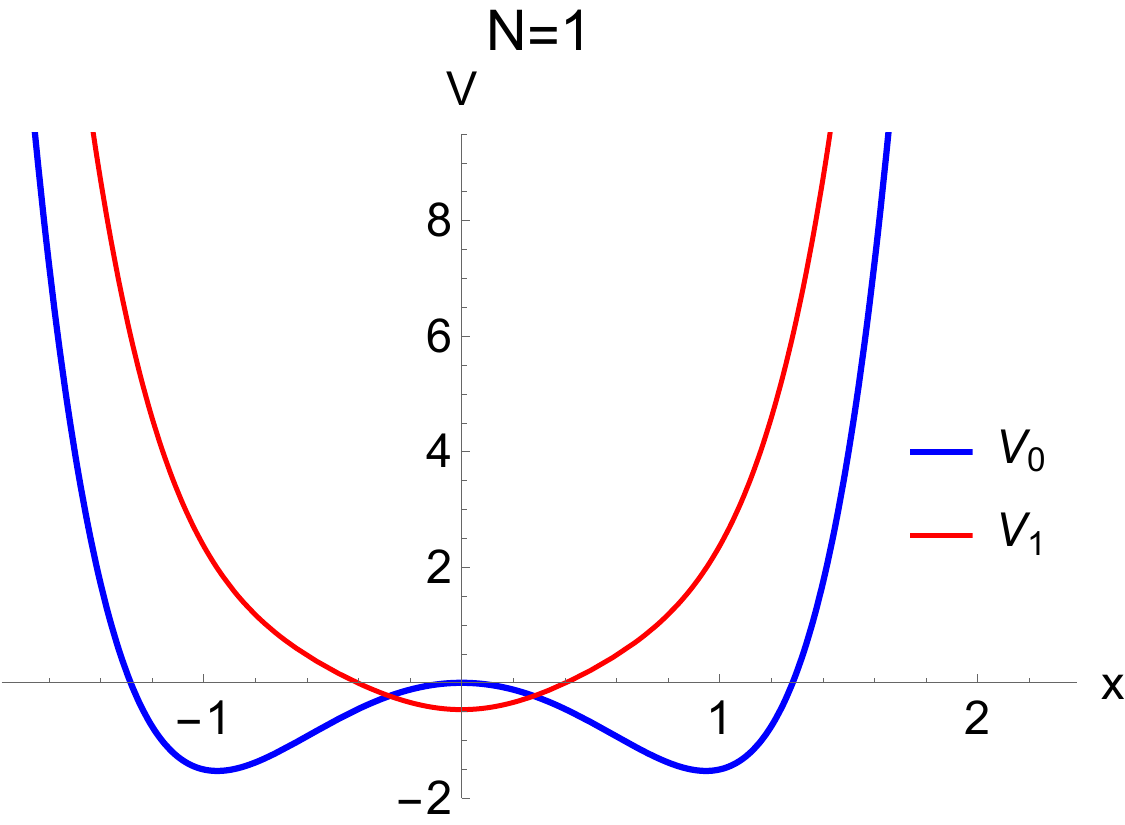} \hspace{0.2cm}
\includegraphics[width=7cm]{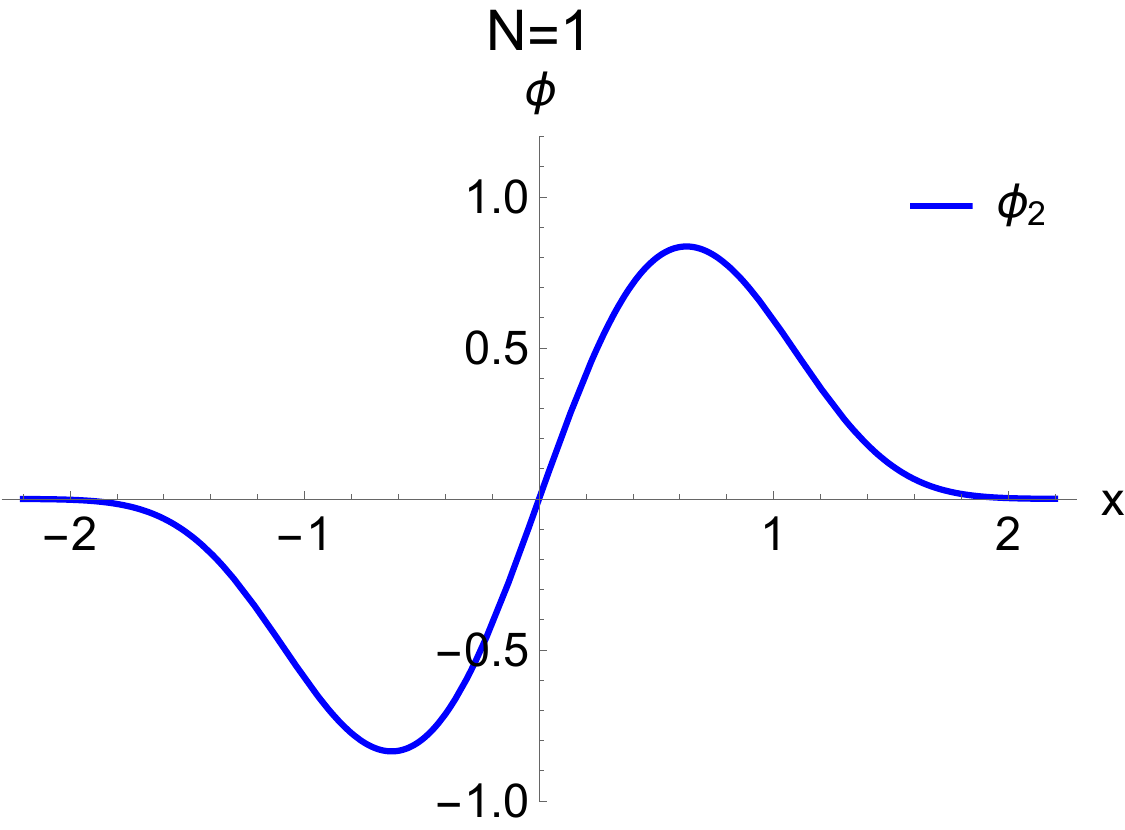}
\caption{Left: For fixed $\mu=\nu=1$, the confining potential $V_0=V^{\rm qes}(x)/2$ in (\ref{vqes}) at $N=1$ (blue curve) and its SUSY partner $V_1$ (red line) as in (\ref{V1 qes}). Right: The (exact) first excited eigenfunction of $\mathcal{H}_{1}$. }
\end{figure}

Now, let us introduce the gauge factor 
\begin{equation}
    \Gamma_1 \ = \  \frac{1}{P^{(N=1)}_0}\,   \exp{[-\frac{1}{4}\,x^{4}-\frac{1}{2}\,x^{2}]}\ ,
\end{equation}
thus, $\phi_2 = x\,\Gamma_1$ and the polynomial factor of $\phi_2$ is simply $x$. Clearly, in the $x$-variable the gauge rotated Hamiltonian $\Gamma_1^{-1}\,{\cal H}_1\, \Gamma_1$ will contain non-polynomial coefficients. To obtain an algebraic differential operator with polynomial coefficients
we construct the gauge-rotated Hamiltonian
\begin{eqnarray}
 \hspace{-2cm} h^{(susy)}|_{N=1} \ & \equiv \ &   2\,P^{(N=1)}_0\,\Gamma_1^{-1}
    \,({\cal H}_1\,-\,\lambda)\, \Gamma_1 \nonumber \\ 
    &  = & \ -(2 x^2+\sqrt{3}+1)\,\frac{d^2}{dx^2} \  + \  2 \,x\, \big[2 x^4+(\sqrt{3}+3) x^2+\sqrt{3}+5\big]\frac{d}{dx} \nonumber \\ 
   & &   - \ \big[ \, 4 x^4-2 \left(\sqrt{3}-2 \lambda \right) x^2-3 \sqrt{3}+1 +2 \left(\sqrt{3}+1\right) \lambda \,\big] \ ,    
\end{eqnarray}
where $\lambda$ is a real constant. That way, the spectral problem $\mathcal{H}_{1}\,\phi=\,E\,\phi$ is equivalent to the zero-mode equation $ h^{(susy)}\,p(x)=0$ being $p=p(x)$ a function to be determined. In this case, $\phi_2(x) = p(x)\,\Gamma_1$ is the solution of the Hamiltonian ${\cal H}_1$.

By construction, at $\lambda=E_2^{(N=1)}= \frac{3}{2}+\sqrt{3}$, the polynomial $p_1=x$ belongs to the kernel of $h^{(susy)}|_{N=1}$, namely $h^{(susy)}|_{N=1}\,p_1=0$. Nevertheless, in the $x$-variable this operator can not be rewritten as a constant coefficient quadratic combination in the $\mathfrak{sl}_2$ generators (\ref{gener}).

\subsection{1-SUSY partner potential $V_1$ with $N=2$}
Here $V_0=\frac{1}{2}V^{\rm qes}(x,N=2)$, for which we know three exact analytical solutions, $\psi_0^{(N=2)}$, $\psi_2^{(N=2)}$ and $\psi_4^{(N=2)}$. Substituting $u=\psi_0$ in (\ref{V1 susy}) we obtain
\begin{eqnarray} \label{V2 qes}
 \hspace{-2cm}   V_1(x,N=2)  & = &  \ \frac{x^6}{2}\ + \ x^4 \ - \ 2 \,x^2\ + \ 1 \ + \ \frac{4 \left(2 x^2-3\right)}{P^{(N=2)}_0}\ + \ \frac{48\, x^2}{\left[P^{(N=2)}_0\right]^2} \nonumber \\ 
&= & \ \frac{1}{2} V^{\rm qes}(x,N=1/2)\  + \ 1 \ + \ \frac{4 \left(2 x^2-3\right)}{P^{(N=2)}_0}\ + \ \frac{48\, x^2}{\left[P^{(N=2)}_0\right]^2}\ ,    
\end{eqnarray}
here $P^{(N=2)}_0=2 x^4+6 x^2+3$. For the above potential $V_1$ we know two exact solutions: a first excited state
\begin{equation}
  \hspace{-2cm}  \phi_2 \ = 
  \  \frac{\big(\left(6 \sqrt{2}-4\right) x^5+\left(8 \sqrt{2}+4\right) x^3+3 \left(\sqrt{2}+4\right) x\big)}{P^{(N=2)}_0}\,   \exp \left(-\frac{1}{4} x^4-\frac{1}{2} x^2\right) \ ,
\end{equation}
with energy $E_2^{(N=2)}= \frac{9}{2}-2 \sqrt{2}$, and a third excited state
\begin{equation}
 \hspace{-2cm}  \phi_4 \ = 
 \frac{\big(\left(6 \sqrt{2}+4\right) x^5+\left(8 \sqrt{2}-4\right) x^3+3 \left(\sqrt{2}-4\right) x\big)}{P^{(N=2)}_0}\,   \exp \left(-\frac{1}{4} x^4-\frac{1}{2} x^2\right) \ ,
\end{equation}
with energy $E_4^{(N=2)}= \frac{9}{2}+2 \sqrt{2}$, respectively.

\begin{figure}[t]
\centering
\includegraphics[width=7cm]{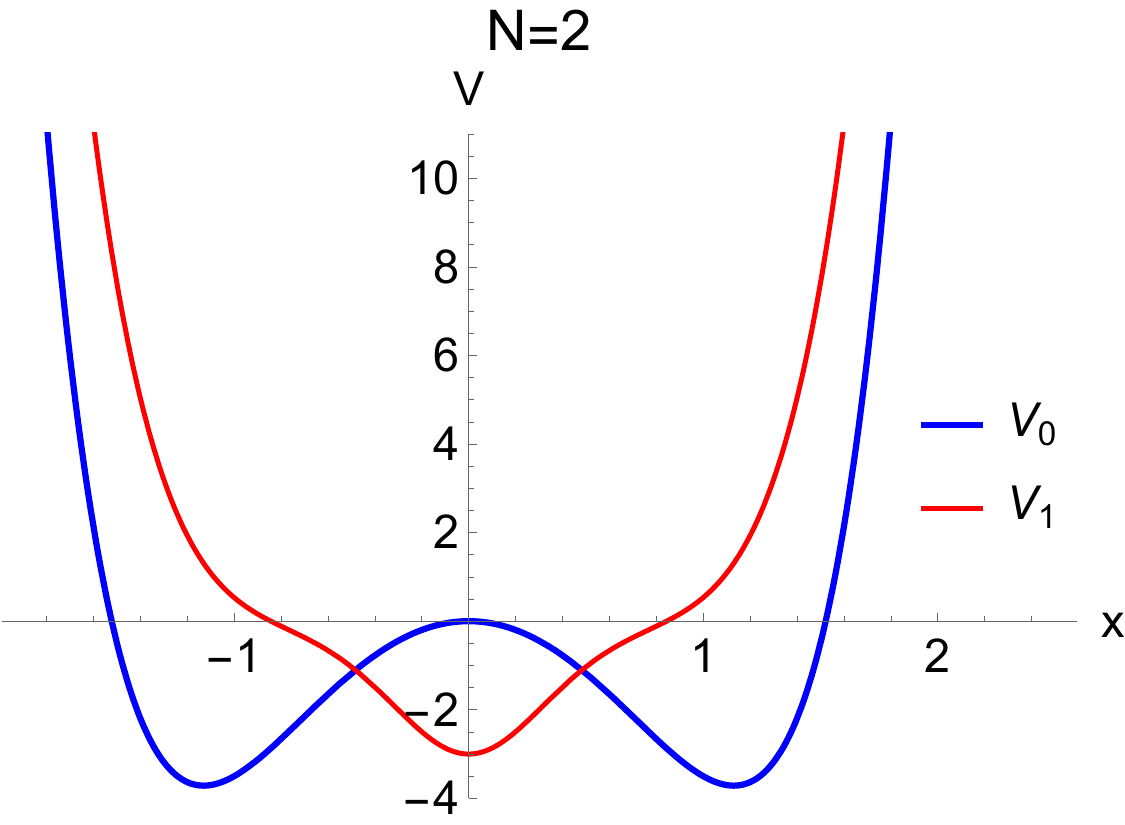} \hspace{0.2cm}
\includegraphics[width=7cm]{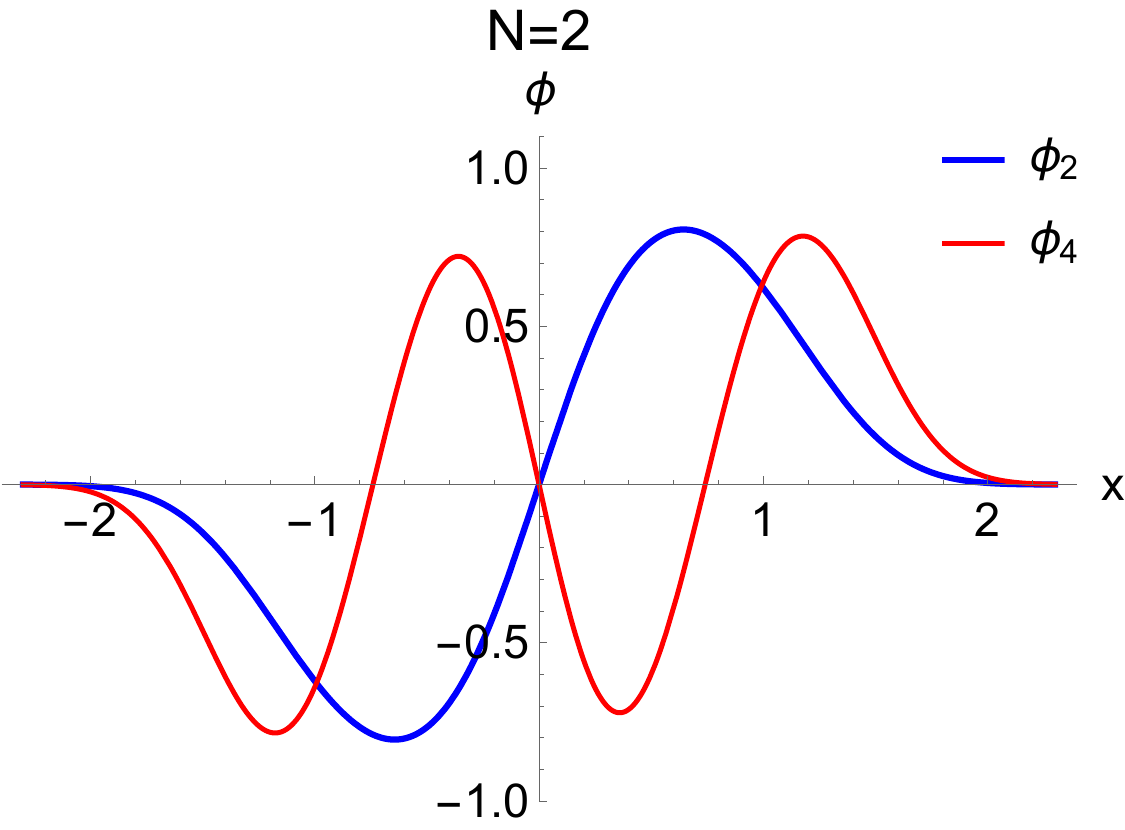}
\caption{Left: For fixed $\mu=\nu=1$, the confining potential $V_0=V^{\rm qes}(x)/2$ in (\ref{vqes}) at $N=2$ (blue curve) and its SUSY partner $V_1$ (red line) as in (\ref{V2 qes}). Right: The (exact) first and third excited eigenfunctions of $\mathcal{H}_{1}$. }
\end{figure}

Now, let us introduce the gauge factor 
\begin{equation}
    \Gamma_2 \ = \  \frac{1}{P^{(N=2)}_0}\,   \exp{[-\frac{1}{4}\,x^{4}-\frac{1}{2}\,x^{2}]}\ .
\end{equation}

Again, in the $x$-variable the gauge rotated Hamiltonian $\Gamma_2^{-1}\,{\cal H}_1\, \Gamma_2$ will contain non-polynomial coefficients. In order to obtain an algebraic differential operator with polynomial coefficients
we build the gauge-rotated Hamiltonian
\begin{eqnarray}
   \hspace{-2cm}  h^{(susy)}|_{N=2} \ & \equiv & \    2\,(2 \,x^4+6 \,x^2+3)\,\Gamma_2^{-2}
    \,({\cal H}_1\,-\,\lambda)\, \Gamma_2  \nonumber 
    \\ 
    & = & \ -\left(2 x^4+6 x^2+3\right)\,\frac{d^2}{dx^2} \  + \  2 \,x\, [2 x^6+8 x^4+17 x^2+15]\frac{d}{dx}  \nonumber \\ 
    & & \ - \ \big[\,20 x^6+(4 \lambda +46) x^4+12 (\lambda +3) x^2+3+6 \lambda  \,\big] \ ,
\end{eqnarray}
where $\lambda$ is a real constant. That way, the spectral problem $\mathcal{H}_{1}\,\phi=\,E\,\phi$ is equivalent to the zero-mode equation $h^{(susy)}\,p=0$. 

By construction, at $\lambda=\frac{9}{2}\pm 2 \sqrt{2}$, the kernel of $h^{(susy)}|_{N=2}$ admits two fifth order polynomials solutions $p_{5,\pm}$, i.e., they obey $h^{(susy)}|_{N=2}\,\,p_{5,\pm}=0$. Nevertheless, in the $x$-variable, this operator can not be rewritten as a constant coefficient quadratic combination in the $\mathfrak{sl}_2$ generators (\ref{gener}).

\subsection{1-SUSY partner potentials $V_1$ with arbitrary integer $N>0$}
Here $V_0=\frac{1}{2}V^{\rm qes}(x,N)$, for which we know $(N+1)$ exact solutions, $\psi_0$, $\psi_2\,\ldots,\psi_{N}$. Substituting $u=\psi_0$ in (\ref{V1 susy}) we arrive to the following expression
\begin{equation} 
\label{Vk qes}
    V_1(x,N)  \
= \ \frac{1}{2} V^{\rm qes}(x,N-3/2)\  + \ 1 \ + \ \frac{Q_{N-1}(x^2)}{P^{(N)}_0}\ + \ \frac{R_{[N/2]}(x^2)}{\left[P^{(N)}_0\right]^2}\ ,
\end{equation}
here $P^{(N)}_0=P^{(N)}_0(x^2)$ is the $2N-$order (in $x-$variable) exact ground state polynomial solution of $h$ (\ref{oph}), whereas $Q_{N-1}$ and $R_{[N/2]}$ are polynomial functions in $x^2$ of order $(N-1)$ and $[N/2]$, respectively \footnote{$[a]$ denotes the integer part of $a$}. For the Hamiltonian $\mathcal{H}_1$ with the above potential $V_1$ (\ref{Vk qes}), the $N$ known exact solutions are given by (\ref{solutions 1susy}). In (\ref{Vk qes}), the part containing no complex poles is given by the sextic potential $\frac{1}{2}V^{\rm qes}(x,N)$ with the replacement $N \rightarrow N-\frac{3}{2}$. 

\vspace{0.2cm}

\textbf{Remark:} \emph{At large distances $|x| \rightarrow \infty$, the last two rational terms in (\ref{Vk qes}) vanish. The dominant term behaves as $\frac{1}{x^2}$. Thus, in this limit the asymptotic behaviour $\phi \sim \exp{[-\frac{1}{4}\,x^{4}-\frac{1}{2}\,x^{2}]}$ of the solutions for $V_1$ is the same as that occurring for $V^{\rm qes}$.}

Using the gauge factor 
\begin{equation}
    \Gamma_N \ = \  \frac{1}{P^{(N)}_0}\,   \exp{[-\frac{1}{4}\,x^{4}-\frac{1}{2}\,x^{2}]}\ ,
\end{equation}
we construct the gauge-rotated Hamiltonian
\begin{eqnarray}
   h^{(susy)}|_{N} \ & \equiv \ &    2\,P^{(N)}_0\,\Gamma_N^{-2}
    \,({\cal H}_1\,-\,\lambda)\, \Gamma_N  \nonumber \\ 
    &  = &  -P^{(N)}_0\,\frac{d^2}{dx^2} \  + \  2 \,x\, Y_{N+1}\,\frac{d}{dx} \ - \ Z_{N+1}  \ ,  
\end{eqnarray}
where $\lambda$ is a real constant and $Y_{N+1}, \,Z_{N+1}$ are polynomial functions in $x^2$ of the same order $(N+1)$. Hence, the spectral problem $\mathcal{H}_{1}\,\phi=\,E\,\phi$ is equivalent to the zero-mode equation $h^{(susy)}|_{N}\,p=0$. 

By construction, the kernel of $h^{(susy)}|_{N}$ admits $N$ polynomial solutions $p_{{}_{4N-3}}$ of order ($4N-3$), i.e., they obey $h^{(susy)}|_{N}\,p_{{}_{4N-3}}=0$. Nevertheless, in the $x$-variable, this operator can not be rewritten as a constant coefficient quadratic combination in the $\mathfrak{sl}_2$ generators (\ref{gener}). 

Moreover, we can recover an intertwining relation for the Hamiltonian $h$ given in (\ref{oph}). From the intertwining relation (\ref{intrel}) we can define the SUSY partner of $h$ as $h_1= \Gamma_N^{-1} h \Gamma_N$ and the intertwining operator as $\mathcal{A}_1^\dag = \Gamma_N^{-1} A_1^\dag \Gamma$, then 
\begin{equation}
    h_1 \,\mathcal{A}_1^\dag \ = \  \mathcal{A}_1^\dag\, h \ . 
\end{equation}
Also, by construction, if $P(z)$ solves $h P(z)=EP(z)$ then $\mathcal{P}(z)= \mathcal{A}_1^\dag P(z)$ solves  $h_1 \mathcal{P}(z)=E\, \mathcal{P}(z)$.

\section{Confluent 2-SUSY generating quasi-exactly solvable potentials}

\subsection{Second-order supersymmetry: confluent case}

Let us present this second-order SUSY transformation as an iteration of two first-order transformations; see more details in \cite{FernandezC.2004,Contreras-Astorga2015a,Contreras-Astorga2015}. Based on the results of the previous subsection \ref{section 1susy}, from $\mathcal{H}_{1}$ we will construct a Hamiltonian $\mathcal{H}_{2}=-\frac{1}{2} \frac{d^2}{dx^2} + V_2(x)$ using a second intertwining operator $A_2^+$, namely
\begin{equation}
    \mathcal{H}_{2} \,A_2^+\ =\ A_2^+\, \mathcal{H}_{1}, \qquad A_2^+ = \frac{1}{\sqrt{2}}\left(- \frac{d}{dx}\ + \ \frac{v'(x)}{v(x)} \right)\ ,
\end{equation}
where the seed function $v(x)$ solves the spectral problem $\mathcal{H}_{1}\, v = \epsilon\, v$. Note that we have used the same factorization energy $\epsilon$ of the previous SUSY step; this is the characteristic feature of the confluent SUSY transformation. The SUSY partner potential becomes 
\begin{equation}
V_2 \ = \ V_1 \ - \ (\ln v)''. \nonumber     
\end{equation}
A simple choice of $v$ comes from (\ref{solutions 1susy}), $v=1/u$; however, this selection leads to $V_2=V_0$. From the reduction-of-order formula, the general choice is given by
\begin{equation}
    v(x)\ = \ \frac{1}{u(x)} \left(\omega_0 \,+\, \int^x u^2(y)\, dy \right) \ ,
\end{equation}
where $\omega_0$ is a constant. To simplify notation, we can define 
\begin{equation}\label{omega confluente}
     \omega(x)\ \equiv \ \omega_0\,+\,\int^x u^2 \,dy \ . 
\end{equation}
Accordingly, the second-order confluent SUSY partner potential of $V_0$ is
\begin{equation} \label{V2 confluent}
    V_2(x) \ =  \ V_0(x) \ - \  \left( \ln \omega(x)\right)'' \ .
\end{equation}
Note that there is an intertwining relation between $\mathcal{H}_{0}$ and $\mathcal{H}_{2}$: $\mathcal{H}_{2} B^+ = B^+ \mathcal{H}_{0}$, where $B^+ = A_2^+ A_1^+$. Moreover, the operator $B^+$ and its adjoint $B= (B^+)^+$ factorize as follows:
\begin{equation}
    B^+ B \,=\, (\mathcal{H}_{2} - \epsilon)^2\ , \qquad B B^+ \,=\, (\mathcal{H}_{0} - \epsilon)^2 \ . 
\end{equation}
The missing state of $\mathcal{H}_{2}$ becomes 
 \begin{equation} \label{missing confluent}
     \phi_\epsilon \, \propto \, \frac{1}{v}\,=\, \frac{u}{\omega(x)} \ , 
 \end{equation}
and the eigenfunctions of $\mathcal{H}_{0}$ are mapped to those of $\mathcal{H}_{2}$ as 
\begin{equation} \label{eigen confluent}
    \phi_n \ = \ \frac{1}{E_n - \epsilon} B^+ \psi_n \quad , \qquad E_n \neq \epsilon \ . 
\end{equation}


Below, we display the confluent SUSY partners of the potentials with energy reflection symmetry presented in subsection \ref{ER symmetry}. To simplify notation, we set $\nu=1$.  

\subsection{Even parity with $N=0$}
To obtain the confluent SUSY partner of $V_0 =\frac{1}{2} V^{\rm qes}_{ER}(x;N=0,\kappa=0)=\frac{1}{2}( x^6\, -\, 3\,x^2)$, besides selecting the seed function which in this case is $u=\psi_{ER}^{N=0}(x; \kappa=0)$ (\ref{ERN0}), it is important to calculate 
\begin{equation}
    \omega(x)\ = \  \omega_0 + \int_0^x u^2(y) \,dy\  = \ \omega_0 + 2^{\frac{1}{4}} \Gamma \left(\frac{5}{4}\right)-\frac{1}{4} \,x \,{\rm E}_{\frac{3}{4}}\left(\frac{x^4}{2}\right)\ , 
\end{equation}
where $\Gamma(z)$ is the Gamma function, and ${\rm E}_n(z)$ is the exponential integral function $E_n(z)=\int^\infty_1 e^{-z t}/t^n ~dt$. We fixed the lower limit of the integral $x_0=0$ to use the parity properties of $u^2$. It is necessary to avoid zeros of $\omega(x)$, for this reason $\omega_0 \in (-\infty, -2^{\frac{1}{4}} \Gamma \left(5/4\right) )\cup (2^{\frac{1}{4}} \Gamma \left(5/4\right), \infty)$ to guarantee the regularity of the potential $V_2= V_0 - (\ln \omega(x))''$.  There is a single analytic eigenfunction (\ref{missing confluent}) of $\mathcal{H}_{2}$
\begin{equation}
    \phi_0 \ = \ \frac{\exp\left(-\frac{1}{4} x^4  \right)}{\omega_0 + 2^{\frac{1}{4}} \Gamma \left(\frac{5}{4}\right)-\frac{1}{4} \,x\, {\rm E}_{\frac{3}{4}}\left(\frac{x^4}{2}\right)}\ ,
    \label{N0confphi0}
\end{equation}
with eigenvalue $\epsilon=0$. 

\begin{figure}[t]
\centering
\includegraphics[width=7cm]{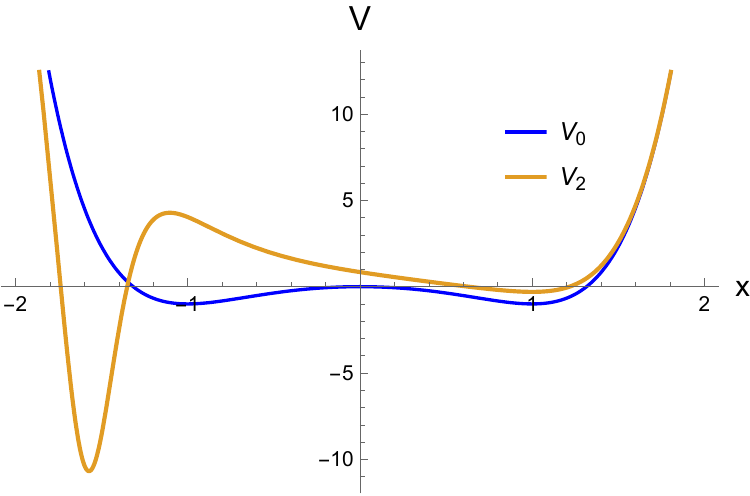}\hspace{0.2cm}
\includegraphics[width=7cm]{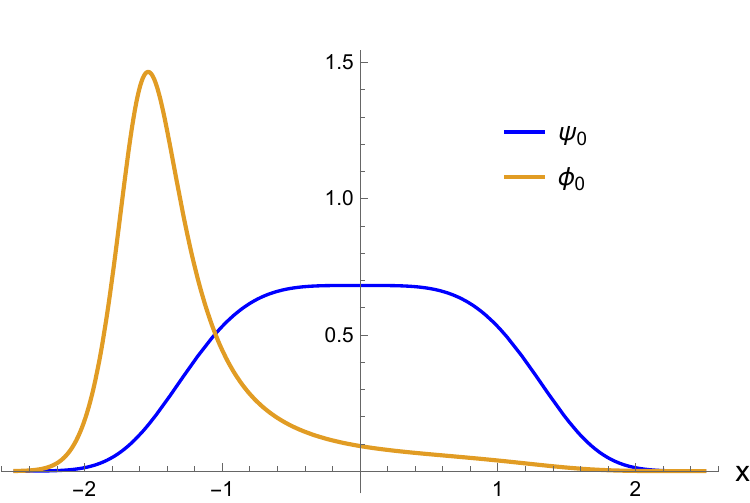}
\caption{Case $N=0$. Left: Plot of the sextic QES potential $V_0 =\frac{1}{2}( x^6\, -\, 3\,x^2)$ with ER symmetry (blue curve) and its confluent SUSY partner (orange curve). Right: Ground state of this sextic potential (blue curve) and the corresponding to the confluent SUSY partner (orange curve). The parameters are $\kappa=0,~\mu = 0,~ \nu = 1,~\omega_0=2^{\frac{1}{4}} \Gamma \left(5/4\right)+0.01$. }
\label{N0confluent}
\end{figure}

The corresponding confluent SUSY partner potential $V_2(x;\kappa=0,N=0)$ and its exact ground state function (\ref{N0confphi0}) are shown in Fig. \ref{N0confluent}. 

\subsection{Odd parity with $N=0$}
For the potential $V_0 =\frac{1}{2} V^{\rm qes}_{ER}(x;N=0,\kappa=1)=\frac{1}{2}( x^6\, -\, 5\,x^2)$, we only have the exact first-excited state. Consequently, using a first-order SUSY transformation will be impossible without adding a singularity to the potential. The confluent-SUSY transformation allows us to create an isospectral Hamiltonian with non-singular potential. In this case $u=\psi_{ER}^{N=0}(x; \kappa=1)=x \exp(- x^4/4)$, see (\ref{ERN0}), then:
\begin{eqnarray}
 \omega(x) & = &  \omega_0 \,+\, \int_0^x u^2(y) \,dy \ = \ \omega_0 + \frac{1}{4} \left(2^{3/4} \Gamma \left(\frac{3}{4}\right)-x^3 E_{\frac{1}{4}}\left(\frac{x^4}{2}\right)\right), \nonumber \\
    \phi_0 &  = & \frac{x \exp(- x^4/4)}{\omega_0 + \frac{1}{4} \left(2^{3/4} \Gamma \left(\frac{3}{4}\right)-x^3 E_{\frac{1}{4}}\left(\frac{x^4}{2}\right)\right)} \ .
\label{N0confphi0odd}       
\end{eqnarray}

\begin{figure}[t]
\centering
\includegraphics[width=7cm]{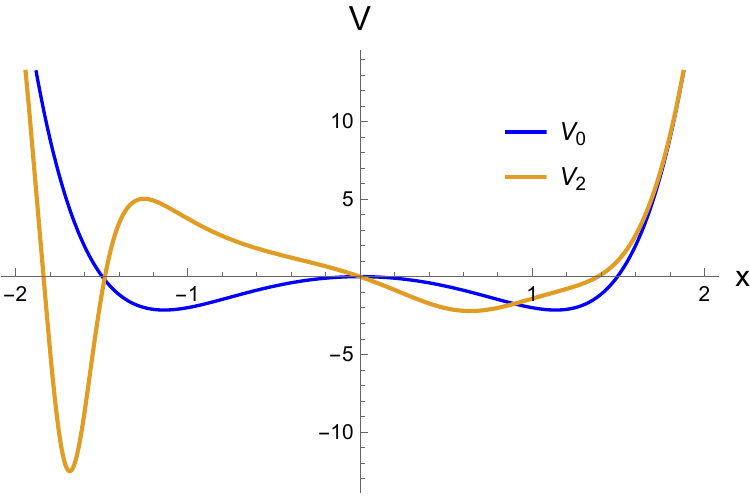}\hspace{0.2cm}
\includegraphics[width=7cm]{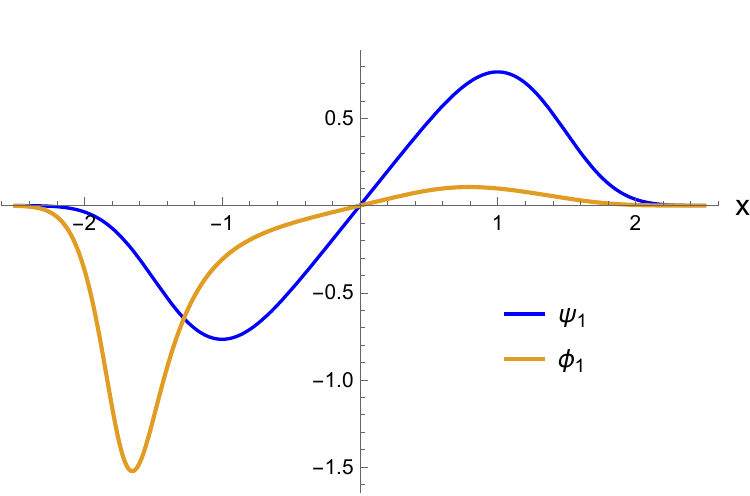}
\caption{Left: Plot of the sextic QES potential $V_0 =\frac{1}{2}( x^6\, -\, 5\,x^2)$ with ER symmetry (blue curve) and its confluent SUSY partner (orange curve). Right: First excited state of $V_0$ (blue curve) and the corresponding of the confluent SUSY partner (orange curve). The parameters are $\kappa=1,~\mu = 0,~ \nu = 1,~\omega_0=0.01 \,+\, 2^{-5/4}~ \Gamma(3/4)$. }
\label{N0confluentodd}
\end{figure}

To avoid singularities on the real line in the potential $V_2=V_0 - (\ln \omega(x))''$, the function $\omega(x)$ must be nodeless, thus  $\omega_0= \left(-\infty,-2^{-5/4}~ \Gamma(3/4)  \right) \cup \left( 2^{-5/4}~ \Gamma(3/4), \infty \right) $. The corresponding confluent SUSY partner potential $V_2(x;\kappa=1,N=0)$ and its exact fist-excited state function (\ref{N0confphi0odd}) are shown in Fig. \ref{N0confluentodd}. 

\textbf{Remark:} \emph{For the confluent SUSY partner potential $V_2(x;\,N=0)$ the original ER symmetry remains. However, the wave functions of the ER symmetric levels are not connected by the original analytical continuation $x \rightarrow i\,x$.} 

\subsection{An example with $N=1$}
Since, at $N=1$, there are two known exact eigenfunctions of $\mathcal{H}_{0}$, there are two different choices for the seed functions $u(x)$. For even parity $\kappa=0$, they are the ground and the second excited states. For odd parity $\kappa=1$, they correspond to the first and the third excited states. The confluent algorithm can be applied as in the previous cases. Firstly, we pick the seed function ($u= \psi^{(N=1)}_{ER,\pm}$, see (\ref{ER case})); so we can construct the function $\omega(x)$ defined as in (\ref{omega confluente}). Secondly, it is important to find the domain of $\omega_0$ such that $\omega(x)$ is nodeless. Thirdly, it is straightforward to obtain the confluent SUSY partner $V_2$ (see (\ref{V2 confluent})). Eventually, there will be two eigenfunctions of $\mathcal{H}_{2}$ associated with the same energies of the initial system $\mathcal{H}_{0}$, one of them $\phi_\epsilon$ is constructed as in (\ref{missing confluent}) and the second one $\phi_n$ is calculated using (\ref{eigen confluent}). As a result, we can build four different Hamiltonians because, for each parity, we have two options for the seed function. 

\subsubsection{Even parity, $u= \psi^{(N=1)}_{ER,-}$}

\begin{figure}[t]
\centering
\includegraphics[width=7cm]{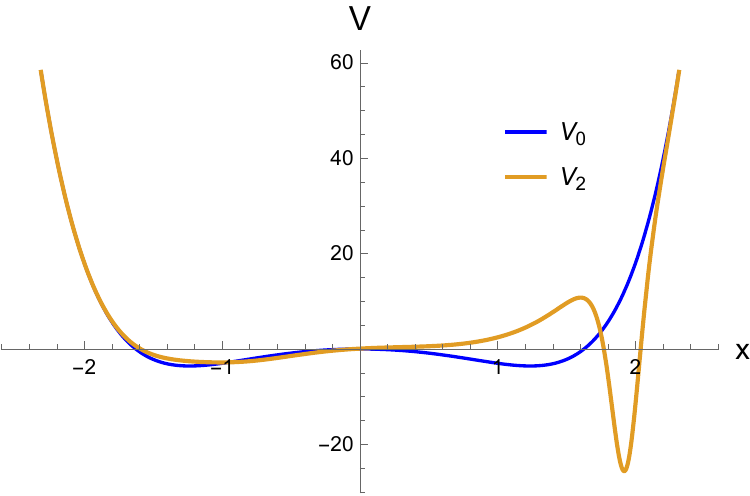}
\caption{Plot of the sextic potential $V_0 =\frac{1}{2}( x^6\, -\, 7\,x^2)$ with ER symmetry (blue curve) and its confluent SUSY partner $V_2$ (orange curve). The parameters are $N=1,~ \kappa=0,~\mu = 0,~ \nu = 1,~\omega_0=-\,0.001-\sqrt[4]{2} \left(\Gamma \left(\frac{1}{4}\right)+2 \Gamma \left(\frac{3}{4}\right)\right) $. }
\label{F7}
\end{figure}

\begin{figure}[t]
\centering
\includegraphics[width=7cm]{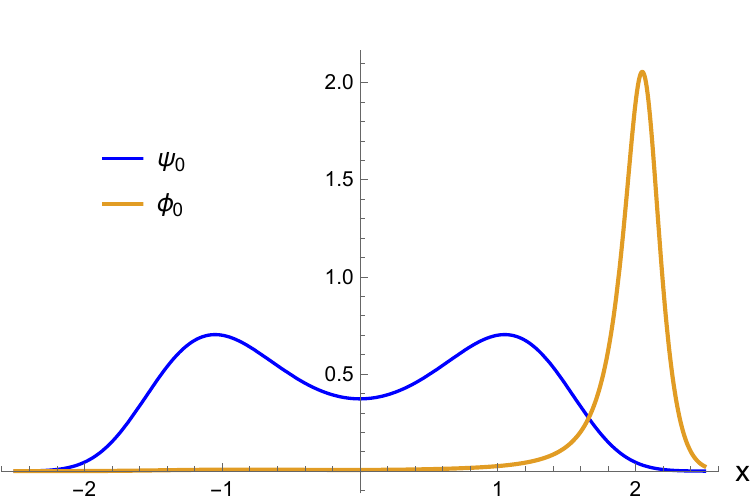}\hspace{0.2cm}
\includegraphics[width=7cm]{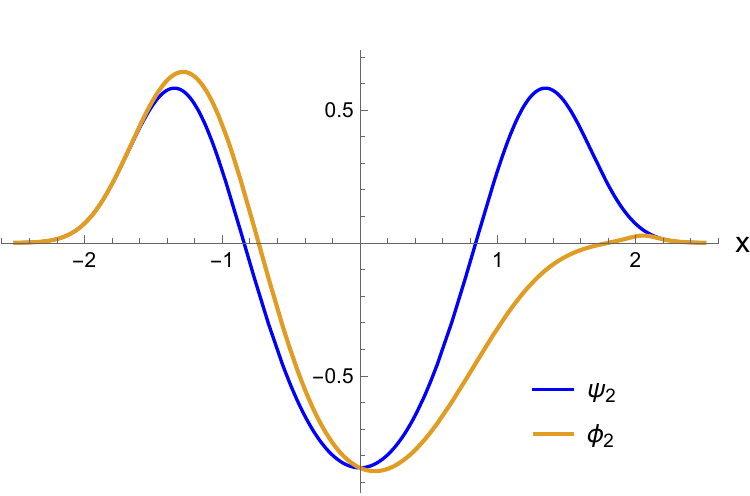}
\caption{Left: Ground state $\psi_0$ of the sextic potential $V_0 =\frac{1}{2}( x^6\, -\, 7\,x^2)$ with ER symmetry (blue curve) and the corresponding solution $\phi_0$ of the
confluent SUSY partner (orange curve). Right: Second excited state $\psi_2$ of $V_0$ (blue curve) and the corresponding solution $\phi_2$ of the confluent SUSY partner (orange curve).  }
\label{F8}
\end{figure}

In this case $u=\psi_{ER,-}^{(N=1)}(x;\kappa=0) \ = \ \big(\,2  \,x^2 + \sqrt{2}\,\big)\, \exp \left(-\frac{1}{4} x^4 \right)$, then:
\begin{eqnarray}
    \hspace{-2cm} \omega(x)& = & \omega_0  -x E_{\frac{3}{4}}\left(\frac{x^4}{2}\right)-\sqrt{2} x^3 E_{\frac{1}{4}}\left(\frac{x^4}{2}\right)-2 e^{-\frac{x^4}{2}} x-\frac{\Gamma \left(-\frac{1}{4}\right)-2 \Gamma \left(\frac{1}{4}\right)}{2^{3/4}} \ , \nonumber \\
    \hspace{-2cm} \phi_0 &  = & \frac{ \big(2\,x^2 +\sqrt{2}\,\big)\, \exp \left(-\frac{1}{4} x^4 \right)}{\omega(x)}                     \ , \qquad \epsilon = -\sqrt{2} \ , \nonumber \\
    \hspace{-2cm} \phi_2 &  = &  \frac{2}{\sqrt{2}+2x^2}  \exp \left(-\frac{x^4}{2} \right)  \left[1-2x^4- \frac{\exp\left(- \frac{x^4}{2} \right) x \left(2+4x^2(\sqrt{2}+x^2) \right)}{\omega(x)}  \right] ,                    
\label{N1confphi0odd}    
\end{eqnarray}
and $E= \sqrt{2}$. To avoid singularities on the real line in the potential $V_2=V_0 - (\ln \omega(x))''$, the function $\omega(x)$ must be nodeless, thus  $\omega_0  \in \left(-\infty  , -\sqrt[4]{2} \left[\Gamma \left(\frac{1}{4}\right)+2 \Gamma \left(\frac{3}{4}\right)\right]  \right) \cup \left( \sqrt[4]{2} \left[\Gamma \left(\frac{1}{4}\right)+2 \Gamma \left(\frac{3}{4}\right)\right]  , \infty \right) $. The corresponding confluent SUSY partner potential $V_2(x;\kappa=0,N=1)$, their exact ground state $\phi_0$  and second excited state $\phi_2$ eigenfunctions (\ref{N1confphi0odd}) are shown in Figs. \ref{F7} and \ref{F8}.

\section{SUSY in the non-algebraic sector of $V^{(\rm qes)}$: case $N=0$ }

Let us consider again the case $N=0$. For the QES sextic potential 
\[
V_0(x) \ = \ \frac{1}{2}V^{\rm qes}(x,N=0) \ = \ x^{6}\ + \ 2\, x^{4}\ - \ 2\, x^{2} \ ,
\]
only the ground state function $\psi_0$ (\ref{psi0N0}) is know analytically. The corresponding 1-SUSY partner potential $V_1(x)$ can be found immediately. Even though no more exact analytical solutions exist other than $\psi_0$, it is important to emphasize that both potentials, $V_0$ and $V_1$, as well as their corresponding (unknown) eigenfunctions and eigenvalues, are still connected by SUSY means.

The set of excited states of $V_0$ can be calculated using numerical and approximate methods. In particular, for one-dimensional systems, a SUSY scheme based on a hierarchy of Hamiltonians has been introduced in previous works to compute the excited states (see \cite{COOPER1995267} and references therein). However, the accuracy of the so obtained solutions is rather limited.

Therefore, it would be worth analyzing
the effect of a SUSY transformation acting on the approximate solutions of $V_0$. This would generate approximate solutions for $V_1$. The accuracy of approximate solutions can be easily estimated using a direct numerical method. Hence, the concrete question we aim to answer can be formulated as follows: \textit{How does the accuracy of approximate solutions change when a SUSY transformation is applied?}.

We adopt the variational approach to compute the first excited states of $V_0(x)$ as a first step. The corresponding trial functions $\psi_{\rm trial}(x)$ are designed on physical grounds and a criterion of simplicity. The accuracy of the obtained solutions, energies, and wave functions is estimated. Afterward, using the operator $A_1^+$, we calculate approximate solutions $\phi = A_1^+\psi_{\rm trial}$ for $V_1$ and determine how the accuracy is modified. 

\subsection{$V_0=\frac{1}{2}V^{\rm qes}(x,N=0)$: approximate variational solutions}

\subsubsection{First excited state}

For the first excited state of $V_0$, we employ the trial function
\begin{equation}
   \psi_{\rm trial}^{\rm (1st)}(x;c_i,k) \ = \  \bigg(\sum_{i=0}^{k}c_i\,x^{2i+1}\bigg)\times\exp{[-\frac{1}{4}\,x^{4}-\frac{1}{2}\,x^{2}]}\ ,
\label{1state}   
\end{equation}
where the $c_i$, ($i=0,1,2,\ldots,k$), are $(k+1)$-variational parameters to be determined by the minimization procedure of the energy functional. As a result of calculations, we can always put $c_0=1$.
The function (\ref{1state}) possesses the following properties:
\begin{itemize}
    \item the orthogonality condition with the exact ground state function is satisfied identically, $\langle\, \psi_{\rm trial}^{\rm (1st)}\, | \psi_{0}^{\rm (exact)}\,\rangle=0$.
    \item it possesses a definite odd parity $\psi_{\rm trial}(-x)=-\,\psi_{\rm trial}(x)$.
    \item the node is correctly located at $x=0$.
    \item the exponential $\sim {\rm exp}{(-\frac{1}{4}\,x^{4}-\frac{1}{2}\,x^{2})}$ factor captures the exact asymptotic behaviour of the state.
\end{itemize}
The corresponding energy functional
\begin{equation}
E_{\rm var}[\psi_{\rm trial}\,;\,c_i] \ = \ \frac{ \langle\, \psi_{\rm trial}^{\rm (1st)}\, |\, {\cal H}\, |\, \psi_{\rm trial}^{\rm (1st)}\,\rangle }{\langle\, \psi_{\rm trial}^{\rm (1st)}\, |\, \psi_{\rm trial}^{\rm (1st)}\,\rangle} \ ,
\label{evar}   
\end{equation}
can be evaluated analytically in terms of Bessel functions. However, we do not present the corresponding (lengthy) expression explicitly.

\begin{figure}[t]
\centering
\includegraphics[width=7cm]{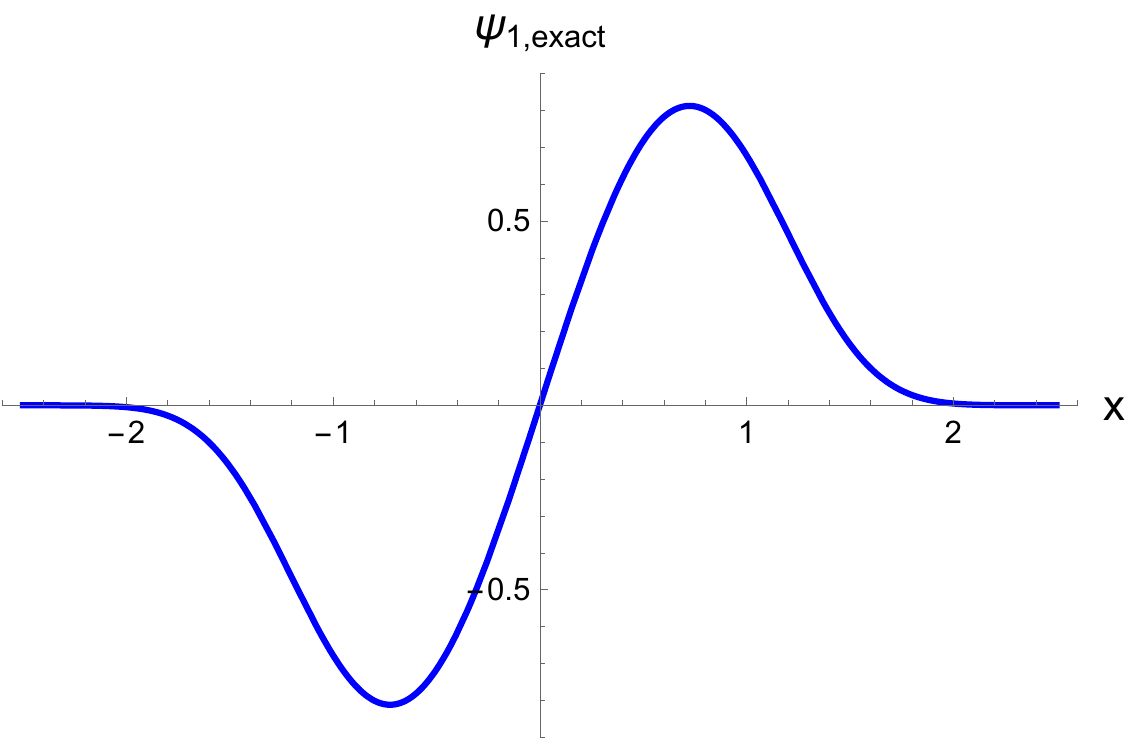} 
\hspace{0.2cm}
\includegraphics[width=7cm]{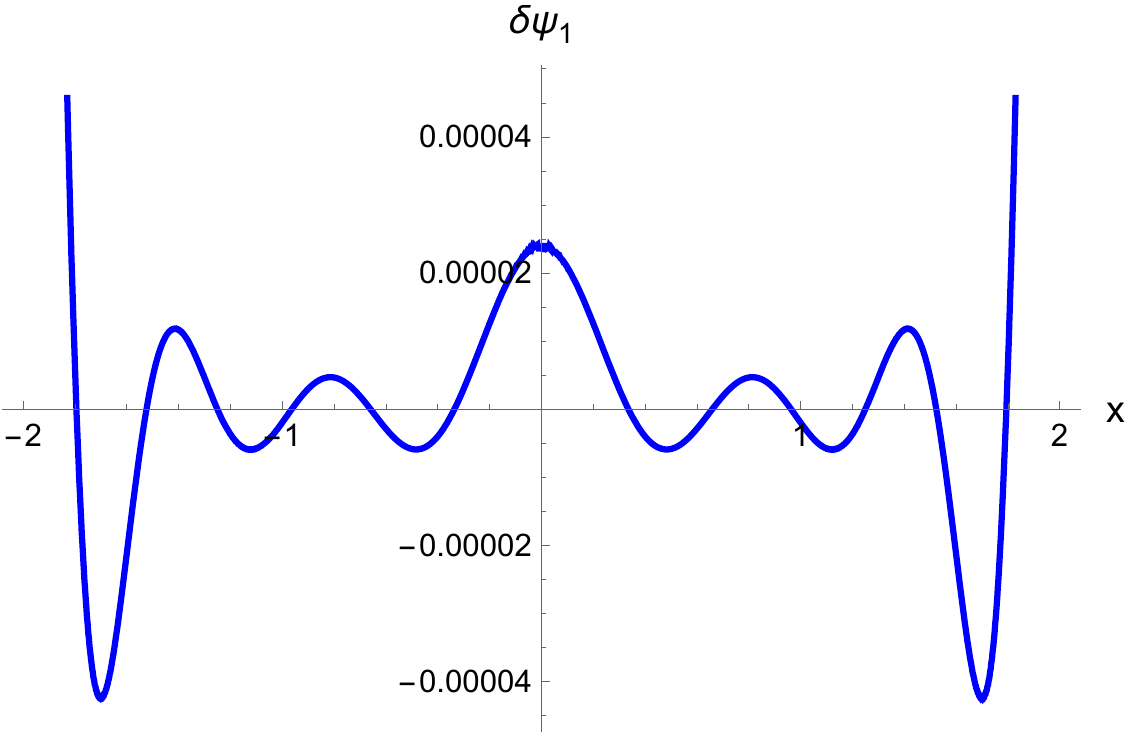}
\caption{First excited state $\psi_1$ of $V_0=\frac{1}{2}V^{\rm qes}(x,N=0)$. The exact numerical wave-function (left), and the local relative error $\delta \psi \equiv \frac{\psi_{\rm trial}-\psi_{\rm exact}}{\psi_{\rm exact}}$ (right) are displayed.}
\label{er1}
\end{figure}

For the lowest values of $k=1,2,\ldots 6$, the minimization of (\ref{evar}) gives the results of the energy $E_{\rm var}^{\rm (1st)}$ displayed in Table \ref{Tab1}. Using the highly accurate LagrangeMesh Mathematica Package\cite{JCR} we obtain the corresponding \textit{exact} numerical result 
\begin{equation}
E_{\rm exact}^{\rm (1st)}\ = \ 2.186500529572814979822675 \ \textrm{a.u.} \ .  \nonumber   
\end{equation}
At $k=6$, the optimal variational parameters are $c_0=1$, 
\begin{eqnarray}
\hspace{-1cm} c_1 \ &=& \ -0.2284993256 \quad ,\quad c_2 \ = \ 
0.06857391226 \quad ,\quad c_3 \ = \ -0.01921167947 \nonumber \\ 
\hspace{-1cm}  c_4 \ &=& \ 0.0042010480 \quad ,\quad c_5 \ = \    -0.0005803713 \quad ,\quad c_6 \ = \     0.0000358148    \nonumber 
\end{eqnarray}
provide a relative error $e_r\equiv \frac{E_{\rm var}-E_{\rm exact}}{E_{\rm exact}}$ of order $\approx 10^{-9}$ with respect to the above exact value $E_{\rm exact}^{\rm (1st)}$. The corresponding relative (local) error $\delta \psi \equiv \frac{\psi_{\rm trial}-\psi_{\rm exact}}{\psi_{\rm exact}}$, of order $10^{-5}$ or less, in the wave-function is shown in Fig. \ref{er1}.
\begin{table}[h]
\centering
\caption{First excited state of $V_0=\frac{1}{2}V^{\rm qes}(x,N=0)$: variational energy obtained using (\ref{1state}) for the lowest values of $k=1,2,3,4,5,6$. The relative error $e_r\equiv \frac{E_{\rm var}-E_{\rm exact}}{E_{\rm exact}}$ is displayed. }
\label{Tab1}
\begin{tabular}{c| c |c}
\hline
\hline
\hspace{0.2cm}  $k$  \hspace{0.4cm}&  \hspace{0.1cm} $E_{\rm var}^{\rm (1st)}$ & $e_r$  \\ \hline
1     & 2.188451041 &  0.000892  \\
2     & 2.186607928 &  0.000049 \\
3     & 2.186506914 &  2.92$\times 10^{-6}$  \\
4     & 2.186500932 &  1.84 $\times 10^{-7}$ \\
5     & 2.186500556 & 1.22$\times 10^{-8}$    \\
6     & 2.186500531  & 8.338$\times 10^{-10}$ \\
\hline
\hline
\end{tabular}
\end{table}

\vspace{0.2cm}

\subsubsection{Second excited state}

Similarly, for the second excited state of $V_0=\frac{1}{2}V^{\rm qes}(x,N=0)$ we use the following trial function

\begin{equation}
   \psi_{\rm trial}^{\rm (2nd)}(x;b_i,k) \ = \  \bigg(\sum_{i=0}^{k}b_i\,x^{2i}\bigg)\times\exp{[-\frac{1}{4}\,x^{4}-\frac{1}{2}\,x^{2}]}\ ,
\label{2state}   
\end{equation}
where the $b_i$, ($i=0,1,2,\ldots,k$), are $(k+1)$-variational parameters to be determined by the minimization procedure of the energy functional and the orthogonality condition. For the values of $k=4,5,6,7$, the energy $E_{\rm var}^{\rm (2nd)}$ is presented in Table \ref{Tab2}.
\begin{table}[h]
\centering
\caption{Second excited state of $V_0=\frac{1}{2}V^{\rm qes}(x,N=0)$: variational energy obtained using (\ref{2state}) for the values of $k=4,5,6,7$.}
\label{Tab2}
\begin{tabular}{c| c |c}
\hline
\hline
\hspace{0.2cm}  $k$  \hspace{0.4cm}&  \hspace{0.1cm} $E_{\rm var}^{\rm (2nd)}$ & $e_r$  \\ \hline
4     & 4.8719031263 &  0.000018 \\
5     & 4.8718230014 &   $1.3\times 10^{-6}$ \\
6     & 4.8718171353 & $9.6\times 10^{-8}  $  \\
7     & 4.8718167005  & $7.3 \times 10^{-9}$ \\
\hline
\hline
\end{tabular}
\end{table}

At $k=7$, the optimal variational parameters are $b_0 \ = \ 1$,
\begin{eqnarray}\label{varpar2}
b_1 \ &=& \  -4.3707510870, \quad b_2 \ = \ 
1.7193375149, \quad b_3 \ = \ -0.6002796514, \nonumber \\ 
b_4 \ &=& \  0.1742370133, \quad b_5 \ = \  -0.0370162648, \quad b_6 \ = \  0.0048080288,  \nonumber \\ 
b_7 \ &=& \  -0.0002756139
\end{eqnarray}
provide the energy $E_{\rm var}^{\rm (2nd)}=4.8718167005$ a.u. with a relative error $e_r\equiv \frac{E_{\rm var}-E_{\rm exact}}{E_{\rm exact}}$ of order $\approx 10^{-8}$ with respect to the exact value 
\begin{equation*}
    E_{\rm exact}^{\rm (2nd)} \ = \ 4.8718166651057818941944\ \textrm{a.u.}\ ,
\end{equation*}
obtained numerically with the LagrangeMesh Mathematica Package\cite{JCR}. The local relative accuracy of $\psi_{\rm trial}^{\rm (2nd)}$ (\ref{2state}) is of order $10^{-4}$ or less. In particular, the two nodes of (\ref{2state}) are located at $x \approx \pm 0.5017$ in agreement with the exact numerical result.  

\begin{figure}[t]
\centering
\includegraphics[width=7cm]{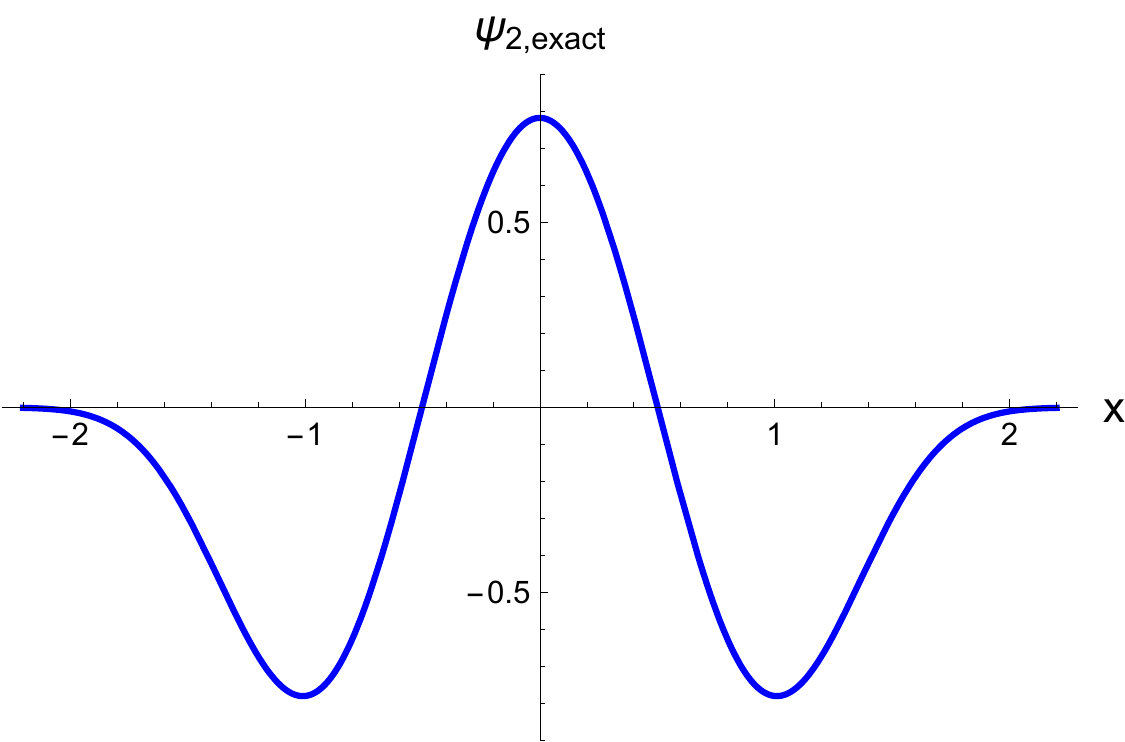} 
\caption{Second excited state $\psi_2$ of $V_0=\frac{1}{2}V^{\rm qes}(x,N=0)$. The exact numerical wave-function is displayed.}
\label{}
\end{figure}

\subsection{Ground state and first excited state of $V_1(x)$: approximate SUSY solutions}

Now, for the 1-SUSY partner potential $V_1$ (\ref{vn0}) we calculate the approximate ground state solution $\phi_{1,\rm trial}$ by simply acting the operator $A_1^+$ (\ref{OpA1}) on the first excited state $\psi_{\rm trial}^{\rm (1st)}$ defined in (\ref{1state}). No further variational minimization is involved. Explicitly,
\begin{equation}
    \phi_{0,\rm trial} \ = \ A_1^+\, \psi_{\rm trial}^{\rm (1st)} \ = \ \bigg(\sum_{i=0}^{k}\,c_i\,(2i+1)\,x^{2i}\bigg)\times\exp{[-\frac{1}{4}\,x^{4}-\frac{1}{2}\,x^{2}]} \ ,
\end{equation}
where the values of parameters $c_i$ correspond to the optimal results obtained for $\psi_{\rm trial}^{\rm (1st)}$ previously. The energy obtained using $\phi_{1,\rm trial}(x)$ as a variational function with no-free parameters is presented in Table \ref{Ephi}.

\begin{table}[h]
\centering
\caption{Ground state energy of $V_1(x; N=0)$: the energy $\epsilon_{1,\rm var}$ is obtained calculating the expectation value of ${\cal H}_1$ on the function $\phi_{1,\rm trial}=A_1^+\, \psi_{\rm trial}^{\rm (1st)}$. The relative error $e_r\equiv \frac{\epsilon_{\rm var}-\epsilon_{\rm exact}}{\epsilon_{\rm exact}}$ is displayed as well.}
\label{Ephi}
\begin{tabular}{c| c |c}
\hline
\hline
\hspace{0.2cm}  $k$  \hspace{0.4cm}&  \hspace{0.1cm} $\epsilon_{1,\rm var}$ & $e_r$  \\ \hline
1     & 2.2043648519 &  0.008  \\
2     & 2.1880809286 &  0.00072 \\
3     & 2.1866346659 &  0.000061  \\
4     & 2.1865117815 &  $5.15 \times 10^{-6}$ \\
5     & 2.1865014728 & $4.31\times 10^{-7}   $ \\
6     & 2.1865006089 &   $3.63\times 10^{-8}$ \\
\hline
\hline
\end{tabular}
\end{table}
\begin{figure}[t]
\centering 
\includegraphics[width=7cm]{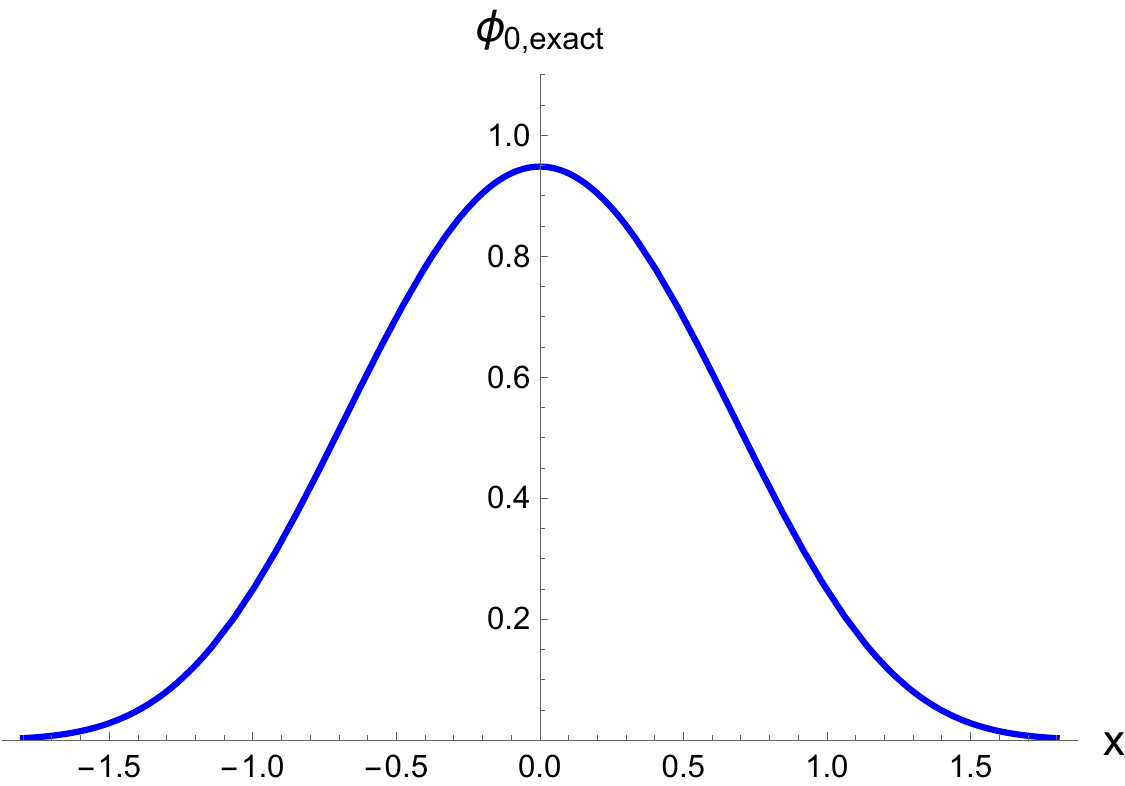} \hspace{0.2cm}
\includegraphics[width=7cm]{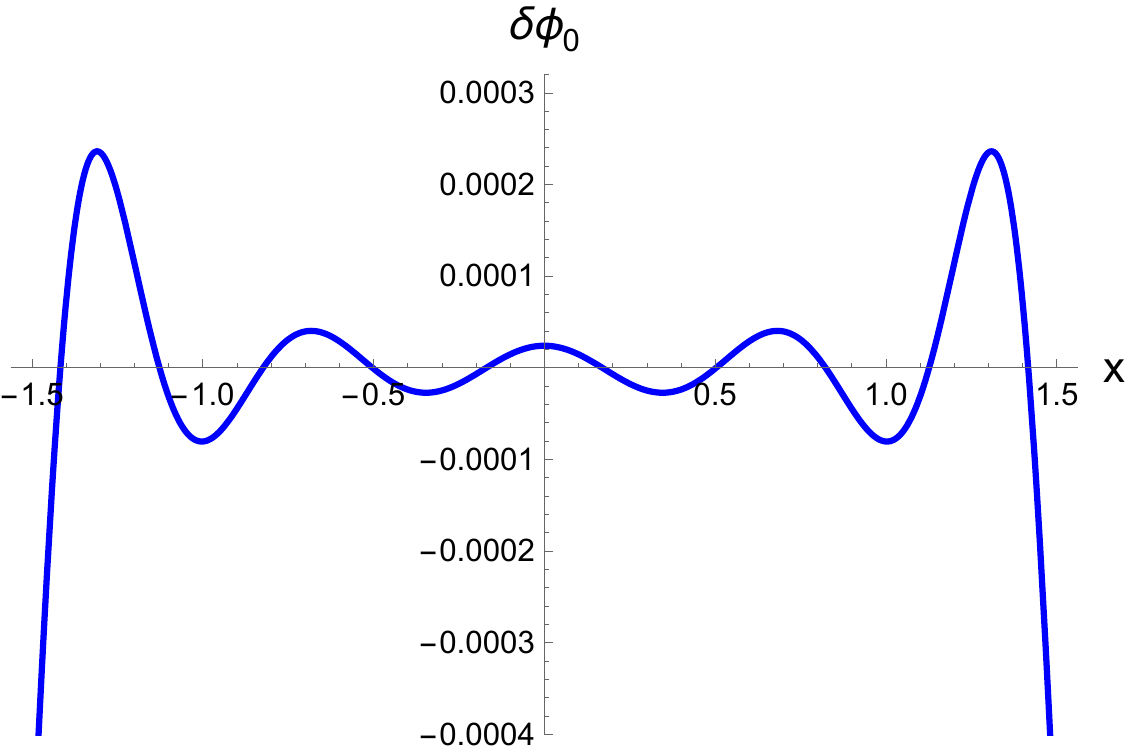}
\caption{Ground state function $\phi_0$ of $V_1(x,N=0)$. The exact numerical wave-function (left), and the local relative error $\delta \phi \equiv \frac{\phi_{\rm trial}-\phi_{\rm exact}}{\phi_{\rm exact}}$ (right) are displayed.  }
\end{figure}

Next, we compute the approximate first excited state of $V_1$ acting the operator $A_1^+$ (\ref{OpA1}) onto the second excited state $\psi_{\rm trial}^{\rm (2nd)}$ defined in (\ref{2state}). Explicitly, 
\begin{equation}
    \phi_{2,\rm trial} \ = \ A_1^+\, \psi_{\rm trial}^{\rm (2nd)} \ = \ \bigg(\sum_{i=1}^{7}\, {2\,i}\,b_i\,x^{2i-1}\bigg)\times\exp{[-\frac{1}{4}\,x^{4}-\frac{1}{2}\,x^{2}]} \ ,
\end{equation}
where the optimal values of parameters $b_i$ are taken from (\ref{varpar2}). In this case the relative error $e_r\equiv \frac{\epsilon_{\rm var}-\epsilon_{\rm exact}}{\epsilon_{\rm exact}}$ is of order $\approx 10^{-7}$ with respect to the exact value $\epsilon_{2}=E_{\rm exact}^{\rm (2nd)}$. By construction, the orthogonality condition $\langle\, \phi_{2,\rm trial}\, | \phi_{0,\rm trial}\,\rangle=0$ is fulfilled exactly.

\begin{figure}[t]
\centering
\includegraphics[width=7cm]{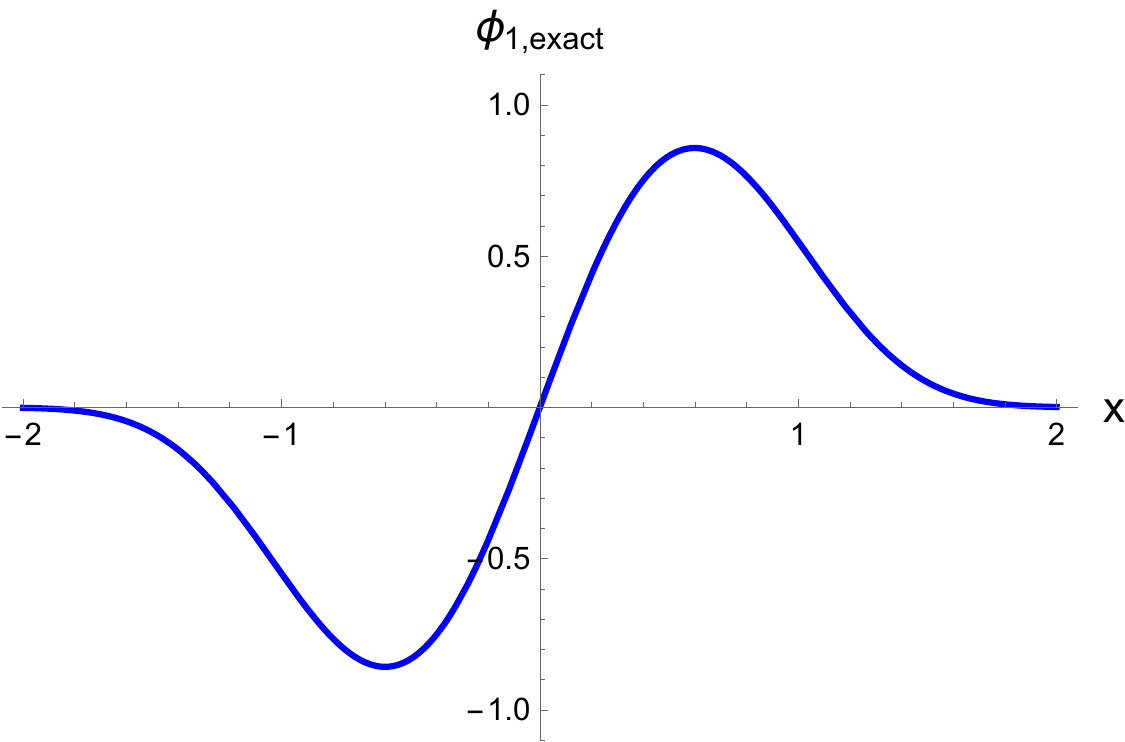} \hspace{0.2cm}
\includegraphics[width=7cm]{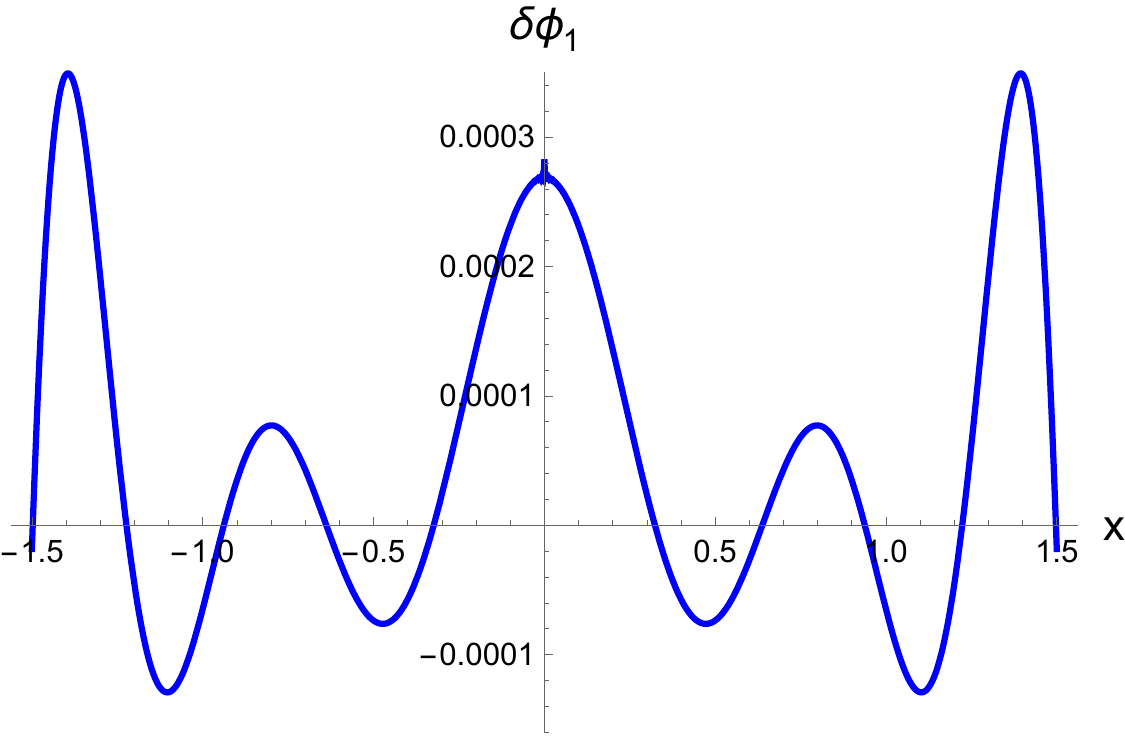}
\caption{First excited state $\phi_1$ of $V_1(x,N=0)$. The exact numerical wave-function (left), and the local relative error $\delta \phi \equiv \frac{\phi_{\rm trial}-\phi_{\rm exact}}{\phi_{\rm exact}}$ (right) are displayed.  }
\end{figure}

\section{Conclusions}

In summary, for the QES sextic potential with integer $N>0$ an algebraic polynomial operator $h(x,\,\partial_x;\,N)$ that governs the $N$ exact polynomial solutions of its 1-SUSY partner $V_1(x)$ is constructed. These odd-parity solutions are polynomials in the variable $x$ of order $(4N-3)$, occurring in the form of zero modes. Nor in the $x$-variable neither in $x^2$, the operator $h(x,\,\partial_x;\,N)$ is $\mathfrak{sl}_2$ Lie-algebraic. In the case $N=0$, the potential $V_1$ possesses a $s\ell_2$ hidden Lie algebra, but no exact solutions occur.

At fixed $N>0$, the potential $V_1(x)$ splits into two additive parts. The first one is polynomial, and it is given by $V^{\rm qes}(x)$ with a different quantized parameter $N\rightarrow N-\frac{3}{2}$, whereas the second part is a rational function in the variable $x^2$. This decomposition represents an important relation between this pair of SUSY partner potentials.    
For instance, at large distances $|x| \rightarrow \infty$ the rational terms in $V_1$ vanish as $\frac{1}{x^2}$. This means that the asymptotic behaviour $\sim \exp{[-\frac{1}{4}\,x^{4}-\frac{1}{2}\,x^{2}]}$ of the solutions for $V_1$ is the same as that occurring for $V^{\rm qes}$. Furthermore, since $V^{\rm qes}(x)$ admits exact analytical solutions at $N=\frac{1}{2},\frac{3}{2},\frac{5}{2},\ldots$,
it follows that for some states with $N>1$ the prefactor $F(x)$ in the solutions of $V_1(x)$, $\phi(x)= F(x)\exp{[-\frac{1}{4}\,x^{4}-\frac{1}{2}\,x^{2}]}$, tends to a polynomial function at large distances $|x| \rightarrow \infty$.  

For the lowest principal quantum numbers $n=0,1,2$ of $V^{(\rm qes)}
$, highly accurate values for the energy ($\sim 20$ s. d.) as a function of $N \in [-1,3]$ were calculated in the non-algebraic sector of the spectrum. In particular, we determined the critical value $N_c=0.732953126$ above which tunneling effects, completely absent in the exact analytical solutions, can occur. As for the first two excited states with $N=0$,  compact physically relevant trial functions are constructed. They are used to estimate the corresponding decrease in accuracy, around one order of magnitude, when supersymmetric quantum mechanics is applied on the level of approximate solutions.

We plan to study interesting open questions such as the possible presence of instanton-like terms in the non-algebraic sector of $V^{(\rm qes)}(x)$, the existence of a hidden Lie algebra of $V_1(x)$ and $V_2(x)$ in a special variable $\tau=\tau(x)$ as well as the corresponding SUSY relations within the framework of path integral, namely, at the level of Feynman diagrams.

\section*{Acknowledgements}
ACA acknowledges Consejo Nacional de Humanidades Ciencia y Tecnolog\'{\i}a (CONAHCyT - M\'exico) support under the grant FORDECYT-PRONACES/61533/2020.

\section*{Data availability}
Data sharing is not applicable to this article as no new data were created or analyzed in this study.

\section*{References}

\bibliographystyle{abbrv}
\bibliography{Biblio}

\end{document}